\documentclass[12pt,epsf]{article}

\usepackage{times}
\usepackage{graphicx}
\usepackage{amsfonts}
\usepackage{amsmath}
\usepackage{amssymb}
\usepackage{amstext}
\usepackage{amsthm}
\usepackage{youngtab}
\usepackage{epsfig}
\usepackage{color}
\psfigdriver{dvips}
\usepackage{latexsym}
\usepackage[matrix,curve]{xypic}
\usepackage{rotating}
\usepackage{slashed}

\rotdriver{dvips}

\begin{document}

\baselineskip 6mm
\renewcommand{\thefootnote}{\fnsymbol{footnote}}


\newcommand{\nc}{\newcommand}
\newcommand{\rnc}{\renewcommand}


\rnc{\baselinestretch}{1.24}    
\setlength{\jot}{6pt}       
\rnc{\arraystretch}{1.24}   

\makeatletter
\rnc{\theequation}{\thesection.\arabic{equation}}
\@addtoreset{equation}{section}
\makeatother



\nc{\be}{\begin{equation}}

\nc{\ee}{\end{equation}}

\nc{\bea}{\begin{eqnarray}}

\nc{\eea}{\end{eqnarray}}

\nc{\xx}{\nonumber\\}

\nc{\ct}{\cite}

\nc{\la}{\label}

\nc{\eq}[1]{(\ref{#1})}

\nc{\newcaption}[1]{\centerline{\parbox{6in}{\caption{#1}}}}

\nc{\fig}[3]{

\begin{figure}
\centerline{\epsfxsize=#1\epsfbox{#2.eps}}
\newcaption{#3. \label{#2}}
\end{figure}
}


\def\CA{{\cal A}}
\def\CC{{\cal C}}
\def\CD{{\cal D}}
\def\CE{{\cal E}}
\def\CF{{\cal F}}
\def\CG{{\cal G}}
\def\CH{{\cal H}}
\def\CK{{\cal K}}
\def\CL{{\cal L}}
\def\CM{{\cal M}}
\def\CN{{\cal N}}
\def\CO{{\cal O}}
\def\CP{{\cal P}}
\def\CR{{\cal R}}
\def\CS{{\cal S}}
\def\CU{{\cal U}}
\def\CV{{\cal V}}
\def\CW{{\cal W}}
\def\CY{{\cal Y}}
\def\CZ{{\cal Z}}


\def\IB{{\hbox{{\rm I}\kern-.2em\hbox{\rm B}}}}
\def\IC{\,\,{\hbox{{\rm I}\kern-.50em\hbox{\bf C}}}}
\def\ID{{\hbox{{\rm I}\kern-.2em\hbox{\rm D}}}}
\def\IF{{\hbox{{\rm I}\kern-.2em\hbox{\rm F}}}}
\def\IH{{\hbox{{\rm I}\kern-.2em\hbox{\rm H}}}}
\def\IN{{\hbox{{\rm I}\kern-.2em\hbox{\rm N}}}}
\def\IP{{\hbox{{\rm I}\kern-.2em\hbox{\rm P}}}}
\def\IR{{\hbox{{\rm I}\kern-.2em\hbox{\rm R}}}}
\def\IZ{{\hbox{{\rm Z}\kern-.39em\hbox{\rm Z}}}}


\def\a{\alpha}
\def\b{\beta}
\def\d{\delta}
\def\ep{\epsilon}
\def\ga{\gamma}
\def\k{\kappa}
\def\l{\lambda}
\def\s{\sigma}
\def\t{\theta}
\def\w{\omega}
\def\G{\Gamma}


\def\half{\frac{1}{2}}
\def\dint#1#2{\int\limits_{#1}^{#2}}
\def\goto{\rightarrow}
\def\para{\parallel}
\def\brac#1{\langle #1 \rangle}
\def\curl{\nabla\times}
\def\div{\nabla\cdot}
\def\p{\partial}


\def\Tr{{\rm Tr}\,}
\def\det{{\rm det}}


\def\vare{\varepsilon}
\def\zbar{\bar{z}}
\def\wbar{\bar{w}}
\def\what#1{\widehat{#1}}


\def\ad{\dot{a}}
\def\bd{\dot{b}}
\def\cd{\dot{c}}
\def\dd{\dot{d}}
\def\so{SO(4)}
\def\bfr{{\bf R}}
\def\bfc{{\bf C}}
\def\bfz{{\bf Z}}

\begin{titlepage}


\hfill\parbox{3.7cm} {{\tt arXiv:1109.6644}}

\vspace{15mm}

\begin{center}
{\Large \bf  An Efficient Representation of Euclidean Gravity I}

\vspace{10mm}

Jungjai Lee ${}^a$\footnote{jjlee@daejin.ac.kr}, John J. Oh${}^b$\footnote{johnoh@nims.re.kr}
and Hyun Seok Yang${}^c$\footnote{hsyang@ewha.ac.kr}
\\[10mm]

${}^a$ {\sl Department of Physics, Daejin University, Pocheon 487-711, Korea}

${}^b$ {\sl National Institute for Mathematical Sciences, Daejeon 305-390, Korea}

${}^c$ {\sl Institute for the Early Universe, Ewha Womans University, Seoul 120-750, Korea}

\end{center}

\thispagestyle{empty}

\vskip1cm


\centerline{\bf ABSTRACT}
\vskip 4mm
\noindent

We explore how the topology of spacetime fabric is encoded into the local structure of
Riemannian metrics using the gauge theory formulation of Euclidean gravity.
In part I, we provide a rigorous mathematical foundation to prove that
a general Einstein manifold arises as the sum of $SU(2)_L$ Yang-Mills instantons
and $SU(2)_R$ anti-instantons where $SU(2)_L$ and $SU(2)_R$ are normal subgroups of
the four-dimensional Lorentz group $Spin(4) = SU(2)_L \times SU(2)_R$.
Our proof relies only on the general properties in four dimensions:
The Lorentz group $Spin(4)$ is isomorphic to $SU(2)_L \times SU(2)_R$ and
the six-dimensional vector space $\Lambda^2 T^*M$ of two-forms splits
canonically into the sum of three-dimensional vector spaces of self-dual and
anti-self-dual two-forms, i.e., $\Lambda^2 T^*M = \Lambda^2_+ \oplus \Lambda^2_-$.
Consolidating these two, it turns out that the splitting of $Spin(4)$ is deeply correlated with the decomposition of two-forms on four-manifold which occupies a central position in the theory
of four-manifolds.
\\



\vspace{1cm}

\today

\end{titlepage}

\renewcommand{\thefootnote}{\arabic{footnote}}
\setcounter{footnote}{0}

\section{Introduction}

Einstein gravity in $d$-dimensional Euclidean space can be formulated as a gauge theory
based on the textbook statement \ct{mtw-book} that spin connections in $d$-dimensions are gauge fields
of Lorentz group $SO(d)$. The Riemann curvature tensor can then be understood as
the field strength of the $SO(d)$ spin connections from the gauge theory point of view.

Let us systematically apply the gauge theory formulation of Einstein gravity
to four-dimensional Riemannian manifolds \ct{opy,ohya}.
We would like to illustrate how our result stated as a Lemma in Section 4 can be derived
by applying only a couple of general properties in four dimensions.
If $M$ is an oriented Riemannian four-manifold,
the structure group acting on orthonormal frames in the tangent space of $M$ is $SO(4)$.
An elementary but crucial fact for us is that the Lorentz group $SO(4)$
is isomorphic to $SU(2)_L \times SU(2)_R/\mathbb{Z}_2$. Let us simply forget about the
$\mathbb{Z}_2$ factor since we are mostly interested in local descriptions (in the level
of Lie algebras). The isomorphism then means that the $SO(4)$ spin connections can be split
into a pair of $SU(2)_L$ and $SU(2)_R$ gauge fields. Accordingly the Riemann curvature tensor
will also be decomposed into a pair of $SU(2)_L$ and $SU(2)_R$ curvature two-forms.

Another significant point comes into our consideration. In four dimensions,
the six-dimensional vector space $\Lambda^2 T^*M$ of two-forms splits
canonically into the sum of three-dimensional vector spaces of self-dual and anti-self-dual
two forms, i.e., $\Lambda^2 T^*M = \Lambda^2_+ \oplus \Lambda^2_-$ \ct{besse,4man-book}.
It turns out that this Hodge decomposition is deeply correlated with the Lie algebra splitting
of $SO(4) = SU(2)_L \times SU(2)_R$.
This can be understood by the isomorphism between the Clifford algebra $\mathbb{C}l(d)$
in $d$-dimensions and the exterior algebra $\Lambda^* M$ of cotangent bundle
$T^*M$ over a $d$-dimensional Riemannian manifold $M$ \ct{spin-book}.
In this correspondence, the chiral operator $\Gamma^{d+1}$ in even dimensions
corresponds to the Hodge star operation $* :\Lambda^k T^*M \to \Lambda^{d-k} T^*M$ in $\Lambda^* M$.
See Eq. \eq{clifford-ext} for the four-dimensional case.
That is, the Clifford map implies that the Lorentz generators $J^{AB} = \frac{1}{4}[\Gamma^A, \Gamma^B]$ in $\mathbb{C}l(4)$ have one-to-one correspondence with the space $\Lambda^2 T^*M$ of two-forms in $\Lambda^* M$. The spinor representation in even dimensions is reducible and its irreducible
representations are defined by the chiral representations whose Lorentz generators are given by
$J^{AB}_\pm \equiv \frac{1}{2}(1 \pm \Gamma^{d+1})J^{AB}$. The splitting of the Lie algebra
$SO(4) = SU(2)_L \times SU(2)_R$ can then be specified by the chiral generators $J^{AB}_\pm$
as $J^{AB}_+ \in SU(2)_L$ and $J^{AB}_- \in SU(2)_R$. Then the Clifford map between
$J^{AB}$ and $\Lambda^2 T^*M$ implies that the chiral splitting of $SO(4) = SU(2)_L \times SU(2)_R$ is isomorphic to the decomposition $\Lambda^2 T^*M = \Lambda^2_+ \oplus \Lambda^2_-$ of two-forms
on a four-manifold which indeed occupies a central position in the Donaldson's theory
of four-manifolds \ct{4man-book}.

Let us now apply the chiral splitting of $SO(4) = SU(2)_L \times SU(2)_R$ and the Hodge decomposition $\Lambda^2 T^*M = \Lambda^2_+ \oplus \Lambda^2_-$ of two-forms together to Riemann curvature
tensors which consist of $SO(4)$-valued two-forms \ct{ohya}.
In this respect, the 't Hooft symbols defined by Eq. \eq{thooft-symbol} take a superb mission
consolidating the Hodge decomposition and the chiral splitting which intertwines the $SU(2)$ group structure with the spacetime structure of self-dual two-forms \ct{opy}.
The Riemann curvature tensor $R_{MNAB}$
consists of $SO(4)$ Lie algebra indices $A, B$ and two-form indices $M, N$ in $\Lambda^2 T^*M$.
First one may apply the chiral splitting of $SO(4) = SU(2)_L \times SU(2)_R$ to yield the result \eq{riemann-su2}. The result leads to a pair $\big(F^{(+)}, F^{(-)} \big)$ of $SU(2)$ field
strengths in $SU(2)_L$ and $SU(2)_R$, respectively. Since $F^{(\pm)}$ are $SU(2)$-valued two-forms,
one can next apply the Hodge decomposition $\Lambda^2 T^*M = \Lambda^2_+ \oplus \Lambda^2_-$ to
yield the results \eq{dec-su2l} and \eq{dec-su2r}. Combining these two decompositions together leads
to the result \eq{dec-riemann}. In the end the Riemann curvature tensor is decomposed into
four types $\{ (+, +), (+, -), (-, +), (-, -) \}$ depending on the types
of $SU(2)$ chiralities \ct{besse}.
After imposing the first Bianchi identity, $R_{AB} \wedge E^B = 0$, we can swap the role of
the indices $A,B$ and $C,D$ in $R_{ABCD} = E_A^M E_B^N R_{MNCD}$,
i.e., $R_{ABCD} = R_{CDAB}$, which leads to the relation \eq{symm-coeff} between the expansion coefficients and an extra constraint \eq{bian-coeff}. Consequently the decomposition \eq{dec-riemann}
of a general Riemann curvature tensor ends in 20 components \ct{ohya}.

After we have realized that the four-dimensional Euclidean gravity can be formulated
as two copies of $SU(2)$ gauge theories, a natural question arises.
What is the Einstein equation from the gauge theory point of view?
An educated guess would be some equations which are linear in $SU(2)$ field strengths
because Riemmann curvature tensors are composed of a pair $\big(F^{(+)}, F^{(-)} \big)$
of $SU(2)$ field strengths. The most natural object linear in the $SU(2)$ field strengths
will be Yang-Mills instantons. The Lemma proven in Section 4 shows that the inference
is pleasingly true.

Recently, in \ct{yayu}, a similar decomposition of Riemann curvature tensors was applied
to 6-dimensional Riemannian manifolds whose holonomy group is $SO(6) \cong SU(4)/\mathbb{Z}_2$.
Using the $SU(4)$ Yang-Mills gauge theory formulation of 6-dimensional
Riemannian manifolds and the six-dimensional 't Hooft symbols
which realize the isomorphism between $SO(6)$ Lorentz algebra and $SU(4)$ Lie algebra,
it was shown in \ct{yayu} that six-dimensional Calabi-Yau manifolds are equivalent to
Hermitian Yang-Mills instantons in $SU(3)$ Yang-Mills gauge theory.
Indeed some of the formulae in this paper are very parallel to six-dimensional ones.

In a series of papers (I \& II), we will introduce this efficient representation
of Euclidean gravity to uncover the topology of spacetime fabric by consolidating
the chiral splitting of $SO(4) = SU(2)_L \times SU(2)_R$ and the Hodge decomposition
$\Lambda^2 T^*M = \Lambda^2_+ \oplus \Lambda^2_-$ of two-forms.
In part I, we will provide a rigorous mathematical foundation for the Lemma proven in \ct{ohya}
stating that an Einstein manifold always arises as the sum of $SU(2)_L$ Yang-Mills instantons
and $SU(2)_R$ anti-instantons.

\begin{figure}[tbp]
\centering
\epsfig{file=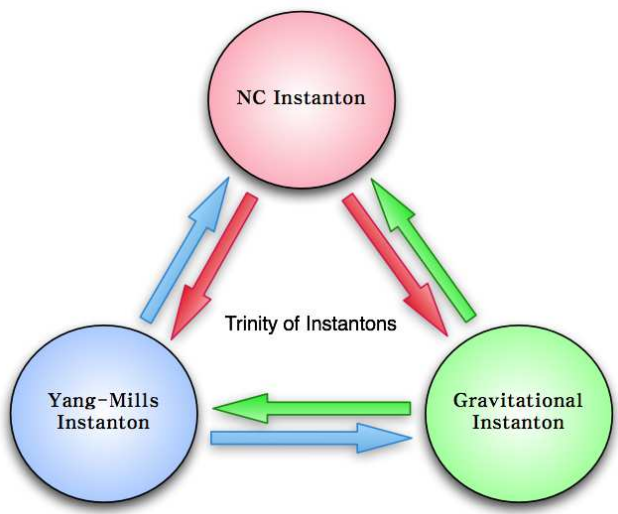,width=0.5\linewidth}
\caption{Trinity of instantons}
\label{fig:trinity}
\end{figure}

The paper is organized as follows. In Section 2, we formulate four-dimensional Euclidean
gravity as $SO(4)$ Yang-Mills gauge theory \ct{opy}.
The explicit relation between gravity and gauge theory
variables will be established. In Section 3, we introduce an irreducible (chiral) spinor
representation of $SO(4)$ which realizes the chiral splitting of $SO(4)$ isomorphic
to $SU(2)_L \times SU(2)_R$. We further show that the chiral splitting of $SO(4)$ is
isomorphic to the Hodge decomposition stating that the six-dimensional vector space
$\Lambda^2 T^*M$ of two-forms splits canonically into the sum of three-dimensional
vector spaces of self-dual and anti-self-dual two-forms, i.e.,  $\Lambda^2 T^*M
= \Lambda^2_+ \oplus \Lambda^2_-$. Consolidating these two, it turns out \ct{ohya}
that the topological classification of four-manifolds is deeply correlated
with the chirality and the self-duality of four-manifolds.
In Section 4, we apply the results in Section 3 to a general Einstein manifold
to uncover what is a corresponding counterpart of the Einstein manifold
from the gauge theory point of view.
We explain a mathematical basis necessary to understand the Lemma in \ct{ohya}.
In Section 5, we survey some geometrical aspects of
K\"ahler manifolds to illustrate the power of our gauge theory formulation and
study the twistor theory of hyper-K\"ahler manifolds. In Section 6, we consider a matter coupling
to see how the energy-momentum tensor of matter fields in the Einstein equations
deforms the structure of an underlying Einstein manifold. The presence of matter fields
in general introduces a mixing of $SU(2)_L$ and $SU(2)_R$ sectors which is absent
in vacuum Einstein manifolds. Finally we address some implications in Section 7 based on
the results obtained in this paper and discuss an intriguing trinity of instantons
shown up in Figure \ref{fig:trinity}. We will conclude with a brief summary of the contents
which will be addressed in the part II \ct{partii}.
An appendix will be devoted to some useful identities of the 't Hooft symbols.

\section{Riemannian Manifolds and Gauge Theory}

Let $M$ be a four-dimensional Riemannian manifold $M$ whose metric is given by
\begin{equation}\label{4-metric}
    ds^2 = g_{MN}(x) dx^M dx^N, \qquad M, N = 1, \cdots, 4.
\end{equation}
Because spinors form a spinor representation of $SO(4)$ Lorentz group which does not arise
from a representation of $GL(4, \mathbb{R})$, in order to couple the spinors to gravity,
it is necessary to introduce at each spacetime point in $M$ a basis of orthonormal
tangent vectors (vierbeins or tetrads) $E_A = E_A^M \partial_M \in \Gamma(TM),
\; A=1, \cdots, 4$ \ct{mtw-book}.
Orthonormality means that $E_A \cdot E_B = \delta_{AB}$. The frame basis $\{ E_A \}$
defines a dual basis $E^A = E^A_M dx^M \in \Gamma(T^*M)$ by a natural pairing
\begin{equation} \label{dual-vector}
\langle E^A, E_B \rangle = \delta^A_B.
\end{equation}
The above pairing leads to the relation $E^A_M E_B^M = \delta^A_B$.
In terms of the non-coordinate (anholonomic) basis in $\Gamma(TM)$ or
$\Gamma(T^*M)$, the metric \eq{4-metric} can be written as
\begin{eqnarray} \label{44-metric}
ds^2 &=& \delta_{AB} E^A \otimes E^B = \delta_{AB} E^A_M E^B_N
\; dx^M \otimes dx^N \nonumber
\\ &\equiv& g_{MN}(x) \; dx^M \otimes dx^N
\end{eqnarray}
or
\begin{eqnarray} \label{inverse-metric}
\Bigl(\frac{\partial}{\partial s}\Bigr)^2 &=& \delta^{AB} E_A \otimes E_B
= \delta^{AB} E_A^M E_B^N \; \partial_M \otimes \partial_N \nonumber
\\ &\equiv& g^{MN}(x)
\; \partial_M \otimes \partial_N.
\end{eqnarray}

There is a large arbitrariness in the choice of a vierbein because the vierbein formalism respects
a local gauge invariance. Under a local Lorentz transformation which is an orthogonal
frame rotation in $SO(4)$, the vectors transform according to
\begin{eqnarray} \label{frame-rotation}
\begin{array}{l}
E_A(x) \to E_A^\prime(x) = E_B(x) {\Lambda^B}_A(x), \\
E^A(x) \to {E^A}^\prime (x) =   {\Lambda^A}_B(x) E^B(x)
\end{array}
\end{eqnarray}
where ${\Lambda^A}_B(x) \in SO(4)$ is a local Lorentz transformation. As in any other discussion of
local gauge invariance, to achieve the local Lorentz invariance requires introducing a gauge field.
On a Riemannian manifold $M$, the spin connection $\omega$ is an $SO(4)$ gauge field \ct{mtw-book}.
To be precise, a matrix-valued spin connection $\omega = \frac{1}{2} \omega_{AB}J^{AB}
= \frac{1}{2} \omega_{MAB}(x) J^{AB} dx^M$ constitutes
a gauge field with respect to the local $SO(4)$ rotations
\begin{equation} \label{spin-so4}
\omega_M \to \omega^\prime_M = \Lambda \omega_M \Lambda^{-1} + \Lambda \partial_M \Lambda^{-1}
\end{equation}
where $\Lambda = \exp(\frac{1}{2} \lambda_{AB}(x)J^{AB}) \in SO(4)$ and
$J^{AB}$ are $SO(4)$ Lorentz generators
which satisfy the following Lorentz algebra
\begin{equation}\label{lorentz-algebra}
    [J^{AB}, J^{CD}] = - \big(\delta^{AC} J^{BD} - \delta^{AD}J^{BC}
    - \delta^{BC}J^{AD} + \delta^{BD}J^{AC} \big).
\end{equation}
Then the covariant derivatives for the vectors in Eq. \eq{frame-rotation} are defined by
\begin{eqnarray} \label{spin-cov}
\begin{array}{l}
D_M E_A = \partial_M E_A - {{\omega_M}^B}_A E_B, \\
D_M E^A = \partial_M E^A + {{\omega_M}^A}_B E^B.
\end{array}
\end{eqnarray}

The connection one-forms ${\omega^A}_B = {{\omega_M}^A}_B dx^M$
satisfy the Cartan's structure equations \cite{mtw-book,egh-report},
\begin{eqnarray} \label{cartan-eq1}
T^A &=& dE^A + {\omega^A}_B \wedge E^B, \\
\label{cartan-eq2}
{R^A}_B &=& d{\omega^A}_B + {\omega^A}_C \wedge {\omega^C}_B,
\end{eqnarray}
where $T^A$ are the torsion two-forms and ${R^A}_B$ are the curvature two-forms.
In terms of local coordinates, they are given by
\begin{eqnarray} \label{cartan-torsion}
&& {T_{MN}}^A = \partial_M E_N^A - \partial_N E_M^A +
{{\omega_M}^A}_B E_N^B -
{{\omega_N}^A}_B E_M^B, \\
\label{cartan-curvature}
&& {{R_{MN}}^A}_B = \partial_M {{\omega_N}^A}_B - \partial_N
{{\omega_M}^A}_B+ {{\omega_M}^A}_C {{\omega_N}^C}_B -
{{\omega_N}^A}_C {{\omega_M}^C}_B.
\end{eqnarray}
Now we impose the torsion free
condition, ${T_{MN}}^A = D_M E_N^A - D_N E_M^A = 0$, to recover the
standard content of general relativity, which eliminates $\omega_M$
as an independent variable, i.e.,
\begin{eqnarray}\label{spin-conn}
\omega_{ABC} &=& E_A^M \omega_{MBC} =  \frac{1}{2} (f_{ABC} - f_{BCA} + f_{CAB}) \nonumber \\
&=& - \omega_{ACB}
\end{eqnarray}
where $f_{ABC}$ are the structure functions defined by
\begin{equation}\label{lie-vector}
    [E_A, E_B] = - {f_{AB}}^C E_C.
\end{equation}
The spin connection \eq{spin-conn} is related to the Levi-Civita connection as follows
\begin{equation}\label{levi-civita}
  {\Gamma_{MN}}^P = {{\omega_M}^A}_B E^P_A E^B_N + E^P_A \partial_M E_N^A,
\end{equation}
which can be derived from the metric-compatibility condition so that the covariant derivative
of the vierbein is zero, i.e.,
\begin{equation}\label{metric-condition}
    D_M E_N^A = \partial_M E^A_N -  {\Gamma_{MN}}^P E_P^A + {{\omega_M}^A}_B E^B_N =0.
\end{equation}

For orthogonal groups the second-rank antisymmetric tensor representation is the same as
the adjoint representation, so the Lorentz generators $J^{AB} = - J^{BA}, \; A,B = 1, \cdots, 4,$
can be conveniently labeled as $T^a, \; a = 1, \cdots, 6$. Hence, we now introduce an
$SO(4)$-valued gauge field defined by $A = A^a T^a$ where $A^a = A^a_{M} dx^M$
are connection one-forms on $M$ and $T^a$ are Lie algebra generators of $SO(4)$ satisfying
\begin{equation}\label{lie-algebra}
    [T^a, T^b] = - f^{abc} T^c.
\end{equation}
The identification \ct{opy,ohya} we want to make is then given by\footnote{It may be
worthwhile to adopt the identification \eq{id} by applying a group homomorphism
of $O(4) = SU(2)_L \times SU(2)_R$. To be precise, the spin connection \eq{id}
is a connection on a spinor bundle induced from the $SO(4)$-bundle and
the structure group of its fiber is lifted to $Spin(4)$, a double cover of $SO(4)$,
according to the short exact sequence of Lie groups: $1 \to \mathbb{Z}_2 \to Spin(4) \to SO(4) \to 1$.
Hence the global isomorphism should refer to $Spin(4)$. Nevertheless we will not care about the $\mathbb{Z}_2$-factor because we are mostly interested in local descriptions (in the level of Lie algebras).}
\begin{equation}\label{id}
    \omega = \frac{1}{2} \omega_{AB} J^{AB} \equiv A = A^a T^a.
\end{equation}
Thereafter, the Lorentz transformation \eq{spin-so4} can be translated into
a usual gauge transformation
\begin{equation}\label{gauge-lorentz-tr}
    A \; \to \; A' = \Lambda A \Lambda^{-1} + \Lambda d \Lambda^{-1}
\end{equation}
where $\Lambda = e^{\lambda^a(x) T^a} \in SO(4)$.
The $SO(4)$-valued Riemann curvature tensor is defined by
\begin{eqnarray} \la{riemann-tensor}
  R &=& d \omega + \omega \wedge \omega \xx
  &=& \frac{1}{2} R_{AB} J^{AB} = \frac{1}{2} \Big(d \omega_{AB}
  + \omega_{AC} \wedge \omega_{CB} \Big) J^{AB} \xx
  &=& \frac{1}{4}\Big(R_{MNAB} J^{AB}\Big) dx^M \wedge dx^N \xx
  &=& \frac{1}{4} \Big[\Big(\partial_M \omega_{NAB} - \partial_N \omega_{MAB}
  + \omega_{MAC} \omega_{NCB} -  \omega_{NAC} \omega_{MCB} \Big) J^{AB} \Big] dx^M \wedge dx^N
\end{eqnarray}
or, in terms of gauge theory variables, it is given by
\begin{eqnarray} \la{curvature-tensor}
  F &=& d A + A \wedge A \xx
  &=& F^a T^a = \Big(d A^a - \frac{1}{2} f^{abc} A^b \wedge A^c \Big) T^a \xx
  &=& \frac{1}{2}\Big(F^a_{MN} T^a\Big) dx^M \wedge dx^N \xx
  &=& \frac{1}{2} \Big[\Big(\partial_M A^a_{N} - \partial_N A^a_{M}
  - f^{abc} A^b_{M} A^c_{N} \Big) T^a\Big] dx^M \wedge dx^N.
\end{eqnarray}
Using the form language where $d = dx^M \partial_M = E^A E_A$
and $A = A_M dx^M = A_A E^A$, the field strength
(\ref{curvature-tensor}) of $SO(4)$ gauge fields in the non-coordinate basis takes the form
\begin{eqnarray} \label{form-field-noncod}
F &=& dA + A \wedge A = \frac{1}{2} F_{AB} E^A \wedge E^B \nonumber \\
&=&  \frac{1}{2} \Big( E_A A_B - E_B A_A + [A_A, A_B] + {f_{AB}}^C A_C \Big) E^A \wedge
E^B
\end{eqnarray}
where we used the structure equation
\begin{equation} \label{structure-eqn}
dE^A = \frac{1}{2} {f_{BC}}^A  E^B \wedge E^C.
\end{equation}

After establishing the identification \eq{id} between gravity and gauge theory variables,
it is straightforward to find a gauge theory representation from formulae
in gravity theory.\footnote{\label{gravity-gauge}Note that it is not always possible. For instance,
the torsion free condition \eq{cartan-eq1} has no counterpart in gauge theory because
the gauge theory has no analogue of vierbeins or tetrads \ct{opy}. Moreover, the converse is not
always true. For example, a Yang-Mills instanton on flat space $\mathbb{R}^4$ does not have
a gravity counterpart because the spin connection on $\mathbb{R}^4$ idetically vanishes.
This issue will be further discussed in the last Section.}
For example, the second Bianchi identity for Riemann curvature tensors is mapped to
the Bianchi identity for Yang-Mills field strengths \ct{opy}, i.e.,
\begin{equation}\label{bianchi-bianchi}
    DR \equiv dR + \omega \wedge R - R \wedge \omega = 0 \quad \Leftrightarrow
    \quad DF \equiv dF + A \wedge F - F \wedge A = 0.
\end{equation}

\section{Spinor Representation and Self-Duality}

In order to make an explicit identification between the spin connections and the
corresponding gauge fields, let us first introduce the four-dimensional Dirac algebra
\begin{equation}\label{dirac}
    \{ \Gamma^A, \Gamma^B \} = 2 \delta^{AB} \mathbf{I}_4,
\end{equation}
where $\Gamma^A \; (A = 1, \cdots, 4)$ are $4$-dimensional Dirac matrices and $\mathbf{I}_n$
denotes an $n \times n$ identity matrix.
Then the $SO(4)$ Lorentz generators are given by
\begin{equation}\label{lorentz-gen}
   J^{AB} = \frac{1}{4} [\Gamma^A, \Gamma^B]
\end{equation}
which satisfy the Lorentz algebra \eq{lorentz-algebra}. It will be useful to have
an explicit representation of Dirac matrices as follows
\begin{equation}\label{dirac-matrix}
    \Gamma^A =  \begin{pmatrix}
                 0 & \sigma^A \\
                 \overline{\sigma}^A & 0
               \end{pmatrix}
\end{equation}
where $\sigma^A = ( i \tau^a, \mathbf{I}_2)$ and $\overline{\sigma}^A = ( -i \tau^a, \mathbf{I}_2)
= (\sigma^A)^\dagger$ and $\tau^a, \; a=1,2,3$ are the Pauli matrices.
Note that the Dirac matrices in Eq. \eq{dirac-matrix} are in the chiral representation where
the chirality matrix $\Gamma^5 \equiv - \Gamma^1 \Gamma^2 \Gamma^3 \Gamma^4$ is given by
\begin{equation}\label{gamma5}
\Gamma^5 =  \begin{pmatrix}
                 \mathbf{I}_2 & 0 \\
                 0 & - \mathbf{I}_2
               \end{pmatrix}.
\end{equation}
The spinor representation of $SO(4)$ is reducible and there are two irreducible
Weyl representations. The Lorentz generators of an irreducible (called Weyl or chiral)
representation are given by
\begin{equation} \la{chiral-gen}
   J^{AB}_\pm = \frac{1}{2} (\mathbf{I}_4 \pm \Gamma^5) J^{AB} \equiv \Gamma_\pm J^{AB}
\end{equation}
where $\Gamma_\pm = \frac{1}{2} (\mathbf{I}_4 \pm \Gamma^5)$.

Consider the product of two Dirac matrices\footnote{Note that
the Dirac matrices defined by \eq{dirac-matrix} are self-adjoint, i.e.,
$(\Gamma^A)^\dagger = \Gamma^A$ and so $\sigma^{AB}$ and $\overline{\sigma}^{AB}$
in Eq. \eq{pro-dirac} are self-adjoint and traceless $2 \times 2$ matrices.
Such a $2 \times 2$ matrix can always be expanded in the basis of the Pauli matrices
which underlies the expansion in Eq. \eq{pro-dirac} and motivates the introduction
of the 't Hooft symbols.}
\begin{equation}\label{pro-dirac}
   \Gamma^A \Gamma^B \equiv \begin{pmatrix}
                 \delta^{AB} \mathbf{I}_2 + i \sigma^{AB}  & 0 \\
                  0 & \delta^{AB} \mathbf{I}_2 + i \overline{\sigma}^{AB}
               \end{pmatrix}
               \equiv \delta^{AB} \mathbf{I}_4 + i \begin{pmatrix}
                 \eta^a_{AB} \tau^a  & 0 \\
                  0 & \overline{\eta}_{AB}^{\dot{a}} \tau^{\dot{a}}
               \end{pmatrix}
\end{equation}
and so the Lorentz generators in Eq. \eq{lorentz-gen} are given by
\begin{equation}\label{logen-thooft}
   J^{AB} = \frac{1}{4} [\Gamma^A, \Gamma^B] = \frac{i}{2} \begin{pmatrix}
                 \eta^{a}_{AB} \tau^a  & 0 \\
                  0 & \overline{\eta}_{AB}^{\dot{a}} \tau^{\dot{a}}
               \end{pmatrix}.
\end{equation}
Here we have distinguished for a later purpose the two kinds of Lie algebra indices
with $a = 1,2,3$ and $\dot{a}=1,2,3$ for $SU(2)_L$ and $SU(2)_R$
in $SO(4) = SU(2)_L \times SU(2)_R$, respectively.
One can see from Eqs. \eq{chiral-gen} and \eq{logen-thooft} that the Lorentz generators
in the positive and negative chirality basis are given by $J^{AB}_+ =
\frac{i}{2} \eta^{a}_{AB} \tau^a$ and $J^{AB}_- = \frac{i}{2} \overline{\eta}_{AB}^{\dot{a}} \tau^{\dot{a}}$, respectively. Thereafter, one can determine two families of $4 \times 4$ matrices,
the so-called 't Hooft symbols \ct{raja-book}, defined by
\begin{equation}\label{thooft-symbol}
\eta^{a}_{AB}  = -i \Tr ( J^{AB}_+ \tau^a), \qquad
\overline{\eta}_{AB}^{\dot{a}} = -i \Tr ( J^{AB}_- \tau^{\dot{a}}).
\end{equation}
An explicit representation of the 't Hooft symbols in the basis \eq{dirac-matrix} is shown up in
Appendix A where we also list some useful identities of the 't Hooft tensors.

One can check that the chiral Lorentz generators $J^{AB}_\pm$ independently satisfy the Lorentz
algebra \eq{lorentz-algebra} from which Eq. \eq{eta-o4-algebra} is deduced and commutes each other, i.e., $[J^{AB}_+, J^{CD}_-] = 0$. They also satisfy the anti-commutation relation
\begin{equation}\label{anti-comm}
    \{J^{AB}_\pm, J^{CD}_\pm\} = -\frac{1}{2} (\delta^{AC} \delta^{BD}
-\delta^{AD}\delta^{BC} \pm \varepsilon^{ABCD}) \Gamma_\pm
\end{equation}
from which Eq. \eq{proj-eta} is deduced. Let us define the right-hand side of Eq. \eq{anti-comm}
as
\begin{equation}\label{4-projection}
P^{ABCD}_\pm \equiv \frac{1}{4} (\delta^{AC} \delta^{BD}
-\delta^{AD}\delta^{BC} \pm \varepsilon^{ABCD}).
\end{equation}
The identity \eq{proj-eta} in turn implies that the above operators can be recapitulated
in an elegant form
\begin{equation}\label{pro-symbol}
 P^{ABCD}_+ = \frac{1}{4} \eta^{a}_{AB} \eta^{a}_{CD}, \qquad  P^{ABCD}_- = \frac{1}{4} \overline{\eta}_{AB}^{\dot{a}}\overline{\eta}_{CD}^{\dot{a}}.
\end{equation}
It is then easy to show that the above operators can serve as a projection operator onto a subspace
of definite chirality, i.e.,
\begin{equation}\label{proj-op}
    P^{ABEF}_\pm P^{EFCD}_\pm = P^{ABCD}_\pm, \qquad P^{ABEF}_\pm P^{EFCD}_\mp = 0.
\end{equation}
Using Eqs. \eq{eta-etabar} and \eq{eta^2}, one can easily derive the following useful properties
\begin{equation}\label{pro-eigen}
    \begin{array}{ll}
      P^{ABCD}_+ \eta^{a}_{CD} = \eta^{a}_{AB}, \qquad & P^{ABCD}_- \eta^{a}_{CD} = 0, \\
      P^{ABCD}_+ \overline{\eta}_{CD}^{\dot{a}} = 0, & P^{ABCD}_- \overline{\eta}_{CD}^{\dot{a}} = \overline{\eta}_{AB}^{\dot{a}},
    \end{array}
\end{equation}
which can be summarized as an important relation \ct{raja-book}
\begin{equation}\label{duality-thooft}
\eta^{a}_{AB} = \frac{1}{2} {\varepsilon_{AB}}^{CD}
\eta^{a}_{CD}, \qquad \overline{\eta}^{\dot{a}}_{AB} = - \frac{1}{2} {\varepsilon_{AB}}^{CD}
\overline{\eta}^{\dot{a}}_{CD}.
\end{equation}

Starting with the chiral representation \eq{dirac-matrix} of the Lorentz algebra,
we have arrived at the self-duality relation \eq{duality-thooft}. In order to closely
understand the interrelation between the chiral representation of Lorentz algebra
and the self-duality, let us introduce the Clifford algebra $\mathbb{C}l(4)$
whose generators are given by
\begin{eqnarray}\label{clifford}
    \mathbb{C}l(4) &=& \{ \mathbf{I}_4, \Gamma^A, \Gamma^{AB}, \Gamma^5 \Gamma^A, \Gamma^5 \} \xx
    &=& \{ \Gamma_+, \Gamma_+^A, \Gamma_+^{AB} \} \oplus \{ \Gamma_-, \Gamma_-^A, \Gamma_-^{AB} \}
\end{eqnarray}
where $\Gamma_\pm^{A} = \Gamma_\pm \Gamma^{A}, \; \Gamma_\pm^{AB} = \Gamma_\pm \Gamma^{AB}$ and $\Gamma^{A_1 A_2 \cdots A_k} = \frac{1}{k!} \Gamma^{[A_1}\Gamma^{A_2} \cdots \Gamma^{A_k]}$
with the complete antisymmetrization of indices. Clifford algebras are closely related to exterior algebras \ct{spin-book}. That is, they are naturally isomorphic as vector spaces.
In fact, the Clifford algebra \eq{clifford} can be identified
with the exterior algebra of a cotangent bundle $T^*M \to M$
\begin{equation}\label{clifford-ext}
    \mathbb{C}l(4) \cong \Lambda^* M = \bigoplus_{k=0}^4 \Lambda^k T^* M
\end{equation}
where the chirality operator $\Gamma^5$ corresponds to the Hodge operator $*: \Lambda^k T^* M \to \Lambda^{4-k} T^* M$.
More precisely, the Clifford algebra may be thought of as a quantization of the exterior algebra,
in the same sense that the Weyl algebra is a quantization of the symmetric algebra \ct{wiki-clifford}.

The spinor representation of $SO(4)$ can be constructed by 2 fermion creation operators
$a^*_1, a^*_2$ and the corresponding annihilation operators $a^1, a^2$ defined
by the gamma matrices in Eq. \eq{dirac-matrix} \ct{group-book}. This fermionic system
can be represented in a four-dimensional Hilbert space $V$ whose states are made by acting
on a Fock vacuum $|\Omega \rangle$, i.e., $a^1|\Omega \rangle = a^2|\Omega \rangle = 0$
with creation operators $a^*_1, a^*_2$, and $a^*_1a^*_2$
\begin{eqnarray}\label{4-fock}
    V &=& |\Omega \rangle \oplus a^*_1|\Omega \rangle \oplus a^*_2|\Omega \rangle
    \oplus a^*_1 a^*_2|\Omega \rangle \xx
    &=& \Big( |\Omega \rangle \oplus a^*_1 a^*_2|\Omega \rangle \Big) \oplus
    \Big( a^*_1|\Omega \rangle \oplus a^*_2|\Omega \rangle \Big).
\end{eqnarray}
Since the chirality operator $\Gamma^5$ commutes with all of the $SO(4)$ Lorentz generators
in Eq. \eq{logen-thooft}, the spinor representation in the Hilbert space $V$ is reducible,
i.e., $V = S_+ \oplus S_-$ and there are two irreducible spinor representations $S_\pm$
each of dimension 2, namely the spinors of positive and negative chirality.
If the Fock vacuum $|\Omega \rangle$ has positive chirality,
the positive chirality spinors of $SO(4)$ are states given by
\begin{equation}\label{p-spinor}
    S_+ = |\Omega \rangle \oplus a^*_1 a^*_2|\Omega \rangle = \mathbf{2}
\end{equation}
while the negative chirality spinors of $SO(4)$ are those obtained by
\begin{equation}\label{n-spinor}
    S_- = a^*_1|\Omega \rangle \oplus a^*_2|\Omega \rangle = \overline{\mathbf{2}}.
\end{equation}
According to the Lie algebra isomorphism $SO(4) = SU(2)_L \times SU(2)_R$,
one may identify two irreducible spinor representations with an $SU(2)_L$ spinor
$\mathbf{2} = S_+$ and an $SU(2)_R$ spinor $\overline{\mathbf{2}} = S_-$.
Because the $SU(2)$ Lie group has only a real representation,
$\overline{\mathbf{2}}$ means not a complex conjugate of $\mathbf{2}$
but a completely independent spinor.

Using the Fierz identity, a tensor product of two spinors in $V$ can be expanded in terms
of the bispinors in Eq. \eq{clifford}. And the Clifford map \eq{clifford-ext} also implies
that a $p$-form $\Psi \in \Lambda^p T^* M$ can be mapped to a bispinor in $\mathbb{C}l(4)$:
\begin{equation}\label{clifford-map}
   \Psi = \frac{1}{p!} \Psi^{(p)}_{A_1 A_2 \cdots A_p} E^{A_1} \wedge E^{A_2} \wedge
   \cdots \wedge E^{A_p} \quad \Leftrightarrow \quad  \slashed{\Psi} =
   \Psi^{(p)}_{A_1 A_2 \cdots A_p} \Gamma^{A_1 A_2 \cdots A_p}.
\end{equation}
Therefore it will be useful to classify the Clifford generators in Eq. \eq{clifford}
in terms of direct products of the Weyl spinors $\mathbf{2}$ and $\overline{\mathbf{2}}$
in Eqs. \eq{p-spinor} and \eq{n-spinor}. The result should be familiar as \ct{group-book}
\begin{eqnarray} \la{22}
&&  \mathbf{2} \otimes \mathbf{2}  = \mathbf{1} \oplus \mathbf{3} =
\{ \Gamma_+, \Gamma_+^{AB} \} = \{ \mathbf{I}_2, i \sigma^{AB} = i  \eta^a_{AB} \tau^a \},  \\
\la{2*2*}
&&  \overline{\mathbf{2}} \otimes \overline{\mathbf{2}} = \overline{\mathbf{1}}
\oplus \overline{\mathbf{3}}=
\{\Gamma_-, \Gamma_-^{AB} \} = \{ \mathbf{I}_2, i \overline{\sigma}^{AB} = i \overline{\eta}_{AB}^{\dot{a}} \tau^{\dot{a}} \}, \\
\la{22*}
&&  \mathbf{2} \otimes \overline{\mathbf{2}}  = \mathbf{4} = \{ \Gamma^{A}_+ \}
= \{ \sigma^{A} \}, \\
\la{2*2-clifford}
&&  \overline{\mathbf{2}} \otimes \mathbf{2}  = \overline{\mathbf{4}} =
\{ \Gamma^{A}_- \} = \{ \overline{\sigma}^{A} \}.
\end{eqnarray}
In particular, $\sigma^{A}$ in $\mathbf{2} \otimes \overline{\mathbf{2}}$ and
$\overline{\sigma}^{A}$ in $\overline{\mathbf{2}} \otimes \mathbf{2}$ are nothing but
a quoternion and a conjugate quoternion, respectively, that maps spinors of one
chirality to the other. A quoternion determines an isomorphism
between the Euclidean space $\mathbb{R}^4$ and the space of bivectors of $\mathbb{C}^2$ where
a point $x^A$ in $\mathbb{R}^4$ is taken to correspond to a quoternion according to
\begin{equation}\label{quoternion}
\mathbb{X}_{\alpha\dot{\alpha}} = x^A \sigma^{A}_{\alpha\dot{\alpha}} \quad \mathrm{or}
\quad \overline{\mathbb{X}}_{\dot{\alpha}\alpha} = x^A \overline{\sigma}^{A}_{\dot{\alpha}\alpha}
\end{equation}
where $\alpha = 1,2 \in \mathbf{2}$ and $\dot{\alpha} = 1,2 \in \overline{\mathbf{2}}$
are doublet indices on $\mathbb{C}^2$. The spinor indices are raised and lowered with
the $SU(2)$-invariant symplectic forms $\epsilon^{\alpha\beta}, \; \epsilon^{\dot{\alpha}\dot{\beta}}$
and their inverses $\epsilon_{\alpha\beta}, \; \epsilon_{\dot{\alpha}\dot{\beta}}$.

The Hodge $*$-operator acts on a vector space $\Lambda^p T^*M$ of $p$-forms
and defines an automorphism of $\Lambda^2 T^*M$ with eigenvalues $\pm 1$.
Therefore, we have the following decomposition
\begin{equation}\label{2-form-dec}
    \Lambda^2 T^*M = \Lambda^2_+ \oplus \Lambda^2_-
\end{equation}
where $\Lambda^2_\pm \equiv P_\pm \Lambda^2 T^*M$ and $P_\pm = \frac{1}{2}(1 \pm *)$.
The eigenspaces $\Lambda^2_+$ and $\Lambda^2_-$ in Eq. \eq{2-form-dec} are called self-dual and anti-self-dual, respectively. If $\Lambda^2_+$ and $\Lambda^2_-$ take values in a vector bundle $E$,
they are called instantons and anti-instantons \ct{4man-book}.
For instance, the Riemann curvature tensor
in Eq. \eq{riemann-tensor} is an $SO(4)$-valued two-form and thus one can define the self-dual
structure according to the decomposition \eq{2-form-dec}. In this case, the eigenspace
$\Lambda^2_+$ or $\Lambda^2_-$ in Eq. \eq{2-form-dec} is called
a gravitational (anti-)instanton \cite{egh-report}.
Now the Clifford map \eq{clifford-map} together with the self-duality relation \eq{duality-thooft} suggests that the eigenspaces $\Lambda^2_+$ and $\Lambda^2_-$
in Eq. \eq{2-form-dec} take values in the tensor products $\mathbf{2}
\otimes \mathbf{2} = \mathbf{3} \oplus \mathbf{1}$ and
$\overline{\mathbf{2}} \otimes \overline{\mathbf{2}} = \overline{\mathbf{3}} \oplus
\overline{\mathbf{1}}$, respectively, with singlets being removed.

In order to elucidate this aspect in depth, let us consider an arbitrary two-form
\begin{equation}\label{two-form}
    F = \frac{1}{2} F_{MN} dx^M \wedge dx^N =  \frac{1}{2} F_{AB} E^A \wedge E^B
    \in \Lambda^2 T^* M
\end{equation}
and introduce the (3+3)-dimensional basis of two-forms in $\Lambda^2 T^*M$ for each chirality of
$SO(4)$ Lorentz algebra \ct{jjl-hsy}
\begin{equation}\label{two-from}
    J_+^a \equiv \frac{1}{2} \eta^a_{AB} E^A \wedge E^B, \qquad
    J_-^{\dot{a}} \equiv \frac{1}{2} \overline{\eta}^{\dot{a}}_{AB} E^A \wedge E^B.
\end{equation}
It is easy to derive the volume forms below using the identities in Appendix A
\begin{equation}\label{vol-id}
\begin{array}{l}
J^a_+ \wedge J^b_+ = 2 \delta^{ab} \sqrt{g} d^4 x, \\
J^{\dot{a}}_- \wedge J^{\dot{b}}_- = - 2 \delta^{\dot{a}\dot{b}} \sqrt{g} d^4 x, \\
J^a_+ \wedge J^{\dot{b}}_- = 0.
\end{array}
\end{equation}
Using Eqs. \eq{4-projection} and \eq{pro-symbol} in  turn,
one can get the following result
\begin{eqnarray} \la{dec-2f}
  F_{AB} &=& (P_+^{ABCD} + P_-^{ABCD})F_{CD} \xx
  &=& f^a_{(+)} \eta^a_{AB}
  + f^{\dot{a}}_{(-)} \overline{\eta}^{\dot{a}}_{AB} \xx
  &\equiv& F^{(+)}_{AB} + F^{(-)}_{AB}
\end{eqnarray}
where $f^a_{(+)} = \frac{1}{4} F_{AB} \eta^a_{AB}$ and $f^{\dot{a}}_{(-)} = \frac{1}{4} F_{AB} \overline{\eta}^{\dot{a}}_{AB}$. In Eq. \eq{dec-2f}, we have introduced self-dual and
anti-self-dual rank-2 tensors defined by
\begin{equation}\label{f-sd-asd}
 F^{(+)}_{AB}  =  f^a_{(+)} \eta^a_{AB}, \qquad  F^{(-)}_{AB} = f^{\dot{a}}_{(-)} \overline{\eta}^{\dot{a}}_{AB}.
\end{equation}
Then Eq. \eq{duality-thooft} immediately leads to the self-duality relation
\begin{equation} \la{sd-asd}
 F^{(+)}_{AB} = \frac{1}{2} {\varepsilon_{AB}}^{CD} F^{(+)}_{CD}, \qquad
  F^{(-)}_{AB} = - \frac{1}{2} {\varepsilon_{AB}}^{CD}  F^{(-)}_{CD}.
\end{equation}
Plugging the result \eq{dec-2f} into Eq. \eq{two-form} leads to the Hodge
decomposition \eq{2-form-dec} for a generic two-form $F$:
 \begin{eqnarray} \la{hodge-2f}
  F &=& F^{(+)} + F^{(-)} \xx
  &=& f^a_{(+)} J^a_{+} + f^{\dot{a}}_{(-)} J^{\dot{a}}_-.
\end{eqnarray}
Therefore one sees that the 't Hooft symbols $\eta^a_{AB}$ and $\overline{\eta}^{\dot{a}}_{AB}$
have a one-to-one correspondence with the spaces $\Lambda^2_+$ and $\Lambda^2_-$
in Eq. \eq{2-form-dec}, respectively. In other words, one can see that
$ F^{(+)} \in \mathbf{3}$ and $ F^{(-)} \in \overline{\mathbf{3}}$.
As a result, if $F$ is a curvature two-form on a vector bundle $E$,
an instanton can be represented by the basis
$\eta^{a}_{AB} \in \mathbf{3}$ and it lives in the positive-chirality space $S_+ = \mathbf{2}$
while an anti-instanton can be represented by the basis
$\overline{\eta}^{\dot{a}}_{AB} \in \overline{\mathbf{3}}$ and
it lives in the negative-chirality space $S_- = \overline{\mathbf{2}}$ \ct{opy,ohya}.

The Clifford map \eq{clifford-ext} implies that the space of two-forms
in exterior algebra $\Lambda^* M$ has a one-to-one correspondence with $SO(4)$
generators in Clifford algebra $\mathbb{C}l(4)$, i.e., $\Lambda^2 T^*M \cong
\Gamma^{AB} \in SO(4)$. Thus the Hodge decomposition \eq{2-form-dec}
in the exterior algebra $\Lambda^* M$ is isomorphic to the Lie algebra decomposition
$SO(4) = SU(2)_L \times SU(2)_R$. Through the Clifford map \eq{clifford-ext},
the splitting of $SO(4)$ is deeply related to the decomposition of the two-forms on four-manifold
which occupies a central position in the Donaldson's theory of four-manifolds \ct{4man-book}.
We want to emphasize that the 't Hooft symbols $\eta^a_{AB}$ and $\overline{\eta}^{\dot{a}}_{AB}$
in this respect take a superb mission consolidating the Hodge decomposition \eq{2-form-dec} and the Lie algebra isomorphism $SO(4) = SU(2)_L \times SU(2)_R$,
which intertwines the group structure of the index $a =1,2,3 \in SU(2)_L$ and
$\dot{a}=1,2,3 \in SU(2)_R$ with the spacetime structure of the two-form indices $A,B$ \ct{raja-book}.
The 't Hooft symbols at the outset have been introduced to define
the chiral decomposition of Lorentz generators in Eq. \eq{chiral-gen} which
concurrently realizes the Lie algebra isomorphism $SO(4) = SU(2)_L \times SU(2)_R$.
But the isomorphism between the Clifford algebra $\mathbb{C}l(4)$ and the exterior algebra
$\Lambda^* M = \bigoplus_{k=0}^4 \Lambda^k T^* M$ also dictates that the Hodge
decomposition \eq{2-form-dec} should be in parallel with the chiral decomposition.
After all, the chirality and the self-duality, which are arguably the most important properties
regarding to the topological classification of Riemannian manifolds \ct{4man-book},
have been amalgamated into the 't Hooft symbols.
A deep geometrical meaning of the 't Hooft symbols is to
specify the triple $(I,J,K)$ of complex structures of a hyper-K\"ahler manifold
for a given orientation. The triple complex structures $(I,J,K)$ form a quaternion which can
be identified with the $SU(2)$ generators $T^a_+$ or $T^{\dot{a}}_-$
in (\ref{thooft-matrix}) \ct{jjl-hsy}.

\section{Einstein Manifolds As Yang-Mills Instantons}

The four dimensional space has mystic features \ct{besse,4man-book}. Among the group of isometries of
$d$-dimensional Euclidean space $\mathbb{R}^d$, the Lie group $SO(4)$
for $d \ge 3$ is the only non-simple Lorentz group and one can define a self-dual two-form
only for $d = 4$. We observed before that these mystic features in four dimensions
can be encoded into the 't Hooft symbols defined by Eq. \eq{thooft-symbol}.
Since the group $SO(4)$ is a direct product of normal subgroups
$SU(2)_L$ and $SU(2)_R$, i.e. $SO(4) = SU(2)_L \times SU(2)_R$,
we take the 4-dimensional defining representation of the Lorentz
generators as follows \ct{opy}
\begin{eqnarray}\label{rep-so4}
    [J^{AB}]_{CD} &=& \frac{1}{2} \Big(\eta^a_{AB} [T^a_+]_{CD} + \overline{\eta}^{\dot{a}}_{AB} [T^{\dot{a}}_-]_{CD} \Big) \xx
    &=& \frac{1}{2} \Big(\eta^a_{AB} \eta^a_{CD} + \overline{\eta}^{\dot{a}}_{AB} \overline{\eta}^{\dot{a}}_{CD} \Big),
\end{eqnarray}
where $T^a_+$ and $T^{\dot{a}}_-$ are the $SU(2)_L$ and $SU(2)_R$ generators given by
Eq. \eq{thooft-matrix}. It is then easy to check using Eqs. \eq{eta-o4-algebra} and
\eq{thooft-su2} that the generators in Eq. \eq{rep-so4} satisfy
the Lorentz algebra \eq{lorentz-algebra}. According to the identification \eq{id},
$SU(2)$ gauge fields can be defined from the spin connections
\begin{eqnarray}\label{spin-gauge}
    [\omega_M]_{CD} &=& \frac{1}{2}\omega_{MAB} [J^{AB}]_{CD} \xx
    &=& \Big( \frac{1}{4} \omega_{MAB} \eta^a_{AB} \Big) [T^a_+]_{CD}
    + \Big( \frac{1}{4} \omega_{MAB} \overline{\eta}^{\dot{a}}_{AB} \Big) [T^{\dot{a}}_-]_{CD} \xx
    &\equiv & A_M^{(+)a} [T^a_+]_{CD} + A_M^{(-)\dot{a}}[T^{\dot{a}}_-]_{CD} = [A_M]_{CD}
\end{eqnarray}
where $A_M^{(+)a}$ and $A_M^{(-)\dot{a}}$ are $SU(2)_L$ and $SU(2)_R$ gauge fields, respectively, defined by
\begin{equation}\label{su2-lr-field}
 A_M^{(+)a} = \frac{1}{4} \omega_{MAB} \eta^a_{AB}, \qquad  A_M^{(-)\dot{a}}
 = \frac{1}{4} \omega_{MAB} \overline{\eta}^{\dot{a}}_{AB}.
\end{equation}
In other words, we get the following decomposition \ct{opy} for spin connections
\begin{equation} \label{spin-sd-asd}
\omega_{MAB} = A_M^{(+)a} \eta^a_{AB} + A_M^{(-)\dot{a}}
\overline{\eta}^{\dot{a}}_{AB}.
\end{equation}
The above decomposition can also be obtained in the exactly same manner as Eq. \eq{dec-2f}.
Plugging Eq. \eq{spin-sd-asd} into Eq. \eq{riemann-tensor} leads to a similar decomposition for
the Riemann curvature tensors
\begin{equation}\label{riemann-su2}
 R_{MNAB} = F^{(+)a}_{MN}\eta^a_{AB} + F^{(-)\dot{a}}_{MN}\overline{\eta}^{\dot{a}}_{AB},
\end{equation}
where
\begin{equation} \label{gi-curvature}
F_{MN}^{(\pm)} = \partial_M A_N^{(\pm)} - \partial_N A_M^{(\pm)} +
[A_M^{(\pm)}, A_N^{(\pm)}].
\end{equation}

Therefore, we see that the four-dimensional Euclidean gravity, when formulated
as the $SO(4)$ gauge theory, will basically be two copies of $SU(2)$ gauge theories \ct{charap-duff}.
Now a natural question arises. If the four-dimensional Euclidean gravity can be formulated
as the $SO(4)$ gauge theory, what is the Einstein equation from the gauge theory point of view?
An educated guess would be some equations which are linear in $SU(2)$ field strengths
because Riemann curvature tensors are composed of a pair of $SU(2)$ field strengths
as was shown in Eq. \eq{riemann-su2}. The most natural object linear in the $SU(2)$ field strengths
will be a Yang-Mills instanton. Now we will recapitulate the following Lemma
proven in \cite{ohya} to show that the inference is true.

$\textbf{Lemma}.$\; If $M$ is an oriented 4-manifold, the spin connections of $M$
are decomposed as Eq. \eq{spin-sd-asd} according to the Lie algebra decomposition
$Spin(4) = SU(2)_L \times SU(2)_R$.
The curvature 2-form can then be written as Eq. \eq{riemann-su2}.
With the decomposition \eq{riemann-su2}, the Einstein equation
\begin{equation} \la{einstein-equation}
R_{AB} - \frac{1}{2}\delta_{AB}R + \delta_{AB}\Lambda=0
\end{equation}
for the 4-manifold $M$ is equivalent to the self-duality equation of Yang-Mills instantons
\begin{equation} \la{ym-instanton-eq}
F^{(\pm)}_{AB} = \pm \frac{1}{2}{\varepsilon_{AB}}^{CD}F^{(\pm)}_{CD},
\end{equation}
where $F^{(+)a}_{AB}\eta^a_{AB} = F^{(-)\dot{a}}_{AB}\overline{\eta}^{\dot{a}}_{AB}=2\Lambda$.

${\it Proof}.$ \; The Hodge $*$-operation is an involution
of $\Lambda^2 T^*M$ which decomposes the two forms into self-dual and anti-self dual parts, $\Lambda^2 T^*M = \Lambda^2_+ \oplus \Lambda^2_-$. Since the field strengths $F^{(\pm)}_{AB}
\equiv E_A^M E_B^N F^{(\pm)}_{MN}$ in Eq. \eq{gi-curvature} consist of $SU(2)$-valued two-forms,
let us apply the Hodge decomposition \eq{dec-2f} to $F^{(\pm)}_{AB}$ to yield \ct{ohya}
\begin{eqnarray}
&& F^{(+)a}_{AB} \equiv f^{ab}_{(++)}\eta^b_{AB} + f^{a\dot{b}}_{(+-)}
\overline{\eta}^{\dot{b}}_{AB}, \la{dec-su2l} \\
&& F^{(-)\dot{a}}_{AB} \equiv f^{\dot{a}b}_{(-+)}\eta^b_{AB} + f^{\dot{a}\dot{b}}_{(--)} \overline{\eta}^{\dot{b}}_{AB}. \la{dec-su2r}
\end{eqnarray}
Using the above result, we get the following decomposition of the Riemann curvature tensor
in Eq. \eq{riemann-su2}
\begin{equation} \la{dec-riemann}
 R_{ABCD}  = f^{ab}_{(++)}\eta^a_{AB} \eta^b_{CD} + f^{a\dot{b}}_{(+-)} \eta^a_{AB} \overline{\eta}^{\dot{b}}_{CD} + f^{\dot{a}b}_{(-+)} \overline{\eta}^{\dot{a}}_{AB} \eta^b_{CD} + f^{\dot{a}\dot{b}}_{(--)} \overline{\eta}^{\dot{a}}_{AB} \overline{\eta}^{\dot{b}}_{CD}.
\end{equation}
Note that the curvature tensors have the symmetry property $R_{ABCD}=R_{CDAB}$
from which one can get the following relations between coefficients
in the expansion \eq{dec-riemann}:
\begin{equation}\label{symm-coeff}
f^{ab}_{(++)} = f^{ba}_{(++)}, \qquad f^{\dot{a}\dot{b}}_{(--)} = f^{\dot{b}\dot{a}}_{(--)},
\qquad f^{a\dot{b}}_{(+-)} = f^{\dot{b}a}_{(-+)}.
\end{equation}
The first Bianchi identity, $\varepsilon^{ACDE}R_{BCDE}=0$, further constrains the coefficients
\begin{equation}\label{bian-coeff}
  f^{ab}_{(++)}\delta^{ab} = f^{\dot{a}\dot{b}}_{(--)}\delta^{\dot{a}\dot{b}}.
\end{equation}

Therefore, the Riemann curvature tensor in Eq. \eq{dec-riemann} has
$20 = (6+6-1) + 9$ independent components, as is well-known \ct{mtw-book}.
The above results can be applied to the Ricci tensor $R_{AB}\equiv R_{ACBC}$ and
the Ricci scalar $R\equiv R_{AA}$ to yield
\begin{eqnarray}  \la{dec-ricci}
 R_{AB} &=& \big(f^{ab}_{(++)} \delta^{ab} + f^{\dot{a}\dot{b}}_{(--)} \delta^{\dot{a}\dot{b}}\big) \delta_{AB} + 2f^{a\dot{b}}_{(+-)}\eta^a_{AC} \overline{\eta}^{\dot{b}}_{BC}, \\
 \la{dec-scalar}
 R &=& 4 \big(f^{ab}_{(++)} \delta^{ab} + f^{\dot{a}\dot{b}}_{(--)} \delta^{\dot{a}\dot{b}}\big),
\end{eqnarray}
where a symmetric expression was taken in spite of the relation \eq{bian-coeff}.
Hence the Einstein tensor $G_{AB} \equiv R_{AB} - \frac{1}{2} R \delta_{AB}$ has 10
independent components given by
\begin{equation}  \la{dec-einstein}
 G_{AB} =  2f^{a\dot{b}}_{(+-)}\eta^a_{AC} \overline{\eta}^{\dot{b}}_{BC} - 2 f^{ab}_{(++)} \delta^{ab} \delta_{AB}.
\end{equation}

A Riemannian manifold satisfying the Einstein equation \eq{einstein-equation},
which can be written as the form $R_{AB} = \Lambda \delta_{AB}$ where $\Lambda$
is a cosmological constant, is called an Einstein manifold.
It is easy to deduce the condition for the Einstein manifold from Eq. \eq{dec-ricci}
which is given by
\begin{equation}\label{cond-einstein}
  f^{ab}_{(++)} \delta^{ab} = f^{\dot{a}\dot{b}}_{(--)} \delta^{\dot{a}\dot{b}}
  = \frac{\Lambda}{2}, \qquad
  f^{a\dot{b}}_{(+-)} = 0.
\end{equation}
Therefore, the curvature tensor for an Einstein manifold reduces to \ct{ohya}
 \begin{eqnarray} \label{einstein-mfd}
 R_{ABCD} &=& F^{(+)a}_{AB}\eta^a_{CD} + F^{(-)\dot{a}}_{AB} \overline{\eta}^{\dot{a}}_{CD} \xx
 &=& f^{ab}_{(++)} \eta^a_{AB} \eta^b_{CD}
 + f^{\dot{a}\dot{b}}_{(--)} \overline{\eta}^{\dot{a}}_{AB} \overline{\eta}^{\dot{b}}_{CD}
\end{eqnarray}
with the coefficients satisfying \eq{cond-einstein}. If $\Lambda =0$,
the result \eq{einstein-mfd} refers to a Ricci-flat manifold.

As was shown in Eq. \eq{sd-asd}, it is obvious that the $SU(2)$ field strengths
in Eq. \eq{einstein-mfd} satisfy the self-duality equation
\begin{equation}\label{su2-inst}
F^{(\pm)}_{AB} = \pm \frac{1}{2} {\varepsilon_{AB}}^{CD} F^{(\pm)}_{CD}.
\end{equation}
And one can easily verify that the converse is true too: If the Riemann curvature tensors are
given by Eq. \eq{einstein-mfd} and so satisfy the self-duality equations \eq{su2-inst},
the Einstein equation \eq{einstein-equation} is automatically satisfied with $2\Lambda = F^{(+)a}_{AB}\eta^a_{AB} = F^{(-)\dot{a}}_{AB}\overline{\eta}^{\dot{a}}_{AB}$.
This completes the proof of the Lemma. $\qquad \qquad \Box$

A few remarks are in order.

The decomposition \eq{dec-riemann} of Riemann curvature tensors can simply be obtained
by applying the projection operators in Eq. \eq{4-projection} to the Riemann tensors:
\begin{equation}\label{proj-riemann}
    R_{ABCD} = (P^{ABA'B'}_+ + P^{ABA'B'}_-) (P^{CDC'D'}_+ + P^{CDC'D'}_-) R_{A'B'C'D'}
\end{equation}
where the coefficients in the expansion \eq{dec-riemann} are given by
\bea \la{coeff++}
&& f^{ab}_{(++)} = \frac{1}{16} \eta^a_{AB} \eta^b_{CD}  R_{ABCD}, \\
\la{coeff--}
&& f^{\dot{a}\dot{b}}_{(--)} = \frac{1}{16} \overline{\eta}^{\dot{a}}_{AB}
\overline{\eta}^{\dot{b}}_{CD}  R_{ABCD}, \\
\la{coeff+-}
&& f^{a \dot{b}}_{(+-)} = \frac{1}{16} \eta^a_{AB} \overline{\eta}^{\dot{b}}_{CD}  R_{ABCD}.
\eea
Therefore, the decomposition \eq{dec-riemann} must be valid for general oriented Riemannian
manifolds although we derived it using the spinor representation of Lorentz algebra.
Actually it can be derived only using the Hodge decomposition \eq{2-form-dec} that is ready
for any oriented four-manifolds and the Lie algebra isomorphism $SO(4) = SU(2)_L \times SU(2)_R$.
Thus the decomposition \eq{dec-riemann} for Riemann curvature tensors is an off-shell statement.
On on-shell, the Einstein equation, $R_{AB} = \Lambda \delta_{AB}$, then enforces no
mixing between $P_+$- and $P_-$-sectors. This mixing can be introduced only through a coupling
to matter fields, as will be shown in Section 6.

It is remarkable to notice that the Bianchi identity \eq{bianchi-bianchi} then guarantees that
every Einstein manifolds which obey Eq. \eq{su2-inst} automatically satisfy the Yang-Mills equation $D_{B}F_{AB} = D^{(+)}_{B}F^{(+)}_{AB} + D^{(-)}_{B}F^{(-)}_{AB} = 0$ \ct{opy}.
This becomes possible because an $SO(4)$-valued quantity can
completely be separated into $SU(2)_L$ and $SU(2)_R$ sectors
according to the Lie algebra isomorphism $SO(4) =SU(2)_L \times SU(2)_R$.
To be precise, the $SO(4)$ field strength is given by $F = F^{(+)} +  F^{(-)} =
F^{(+)a} T^a_+ + F^{(-)\dot{a}} T^{\dot{a}}_-$ where $F^{(\pm)} = dA^{(\pm)} + A^{(\pm)}
\wedge A^{(\pm)}$. The integrability condition, i.e. the Bianchi identity, then reads as
$D^{(\pm)}F^{(\pm)} \equiv dF^{(\pm)} + A^{(\pm)} \wedge F^{(\pm)} - F^{(\pm)} \wedge A^{(\pm)} = 0$
or $\varepsilon^{ABCD} D^{(+)}_{B} F^{(+)}_{CD} = \varepsilon^{ABCD} D^{(-)}_{B} F^{(-)}_{CD} = 0$.
After all, the self-duality equation \eq{su2-inst} leads to $D^{(+)}_{B}F^{(+)}_{AB} = D^{(-)}_{B}F^{(-)}_{AB} = 0$. Therefore, our lemma sheds light on why the action of Einstein gravity is linear in curvature tensors contrary to the Yang-Mills action being quadratic in curvatures.

The trace-free part of the Riemann curvature tensor is called the Weyl tensor \ct{mtw-book}
defined by
\begin{equation}\label{weyl}
W_{ABCD} = R_{ABCD} - \frac{1}{2} \big(\delta_{AC} R_{BD} - \delta_{AD} R_{BC}
- \delta_{BC} R_{AD}  + \delta_{BD} R_{AC} \big) + \frac{1}{6} (\delta_{AC} \delta_{BD}- \delta_{AD} \delta_{BC}) R.
\end{equation}
The Weyl tensor satisfies all the symmetries of the curvature tensor and
all its traces with the metric are zero. Therefore, one can introduce a similar
decomposition for the Weyl tensor
\begin{equation} \la{dec-weyl}
 W_{ABCD}  \equiv g^{ab}_{(++)}\eta^a_{AB} \eta^b_{CD} + g^{a\dot{b}}_{(+-)} \eta^a_{AB} \overline{\eta}^{\dot{b}}_{CD} + g^{\dot{a}b}_{(-+)} \overline{\eta}^{\dot{a}}_{AB} \eta^b_{CD} + g^{\dot{a}\dot{b}}_{(--)} \overline{\eta}^{\dot{a}}_{AB} \overline{\eta}^{\dot{b}}_{CD}.
\end{equation}
The symmetry property of the coefficients in the expansion \eq{dec-weyl}
is the same as Eq. \eq{symm-coeff}
and the traceless condition, i.e. $W_{AB} \equiv W_{ACBC} = 0$,
leads to the constraint for the coefficients:
\begin{equation}\label{cond-weyl}
  g^{ab}_{(++)} \delta^{ab} = g^{\dot{a}\dot{b}}_{(--)} \delta^{\dot{a}\dot{b}} = 0, \qquad
  g^{a\dot{b}}_{(+-)} = g^{\dot{b}a}_{(-+)}= 0.
\end{equation}
Hence the $O(4)$-decomposition for the Weyl tensor is finally given by \ct{ohya}
\begin{equation} \la{final-weyl}
 W_{ABCD}  = g^{ab}_{(++)}\eta^a_{AB} \eta^b_{CD} + g^{\dot{a}\dot{b}}_{(--)} \overline{\eta}^{\dot{a}}_{AB} \overline{\eta}^{\dot{b}}_{CD}
\end{equation}
with the coefficients satisfying \eq{cond-weyl}. One can see that the Weyl tensor has only
$10 = 5+5$ independent components.

By substituting the results \eq{dec-riemann} and \eq{dec-ricci} into Eq. \eq{weyl},
it is straightforward to determine the coefficients $g^{ab}_{(++)} = \frac{1}{16}
\eta^a_{AB} \eta^b_{CD}  W_{ABCD}$ and $g^{\dot{a}\dot{b}}_{(--)} = \frac{1}{16}
\overline{\eta}^{\dot{a}}_{AB} \overline{\eta}^{\dot{b}}_{CD}  W_{ABCD}$ in Eq. \eq{final-weyl}
in terms of the coefficients in curvature tensors:
\begin{eqnarray*}
&& g^{ab}_{(++)} =  f^{ab}_{(++)} - \frac{1}{3} \delta^{ab} f^{cd}_{(++)} \delta^{cd}, \\
&& g^{\dot{a}\dot{b}}_{(--)}  = f^{\dot{a}\dot{b}}_{(--)} - \frac{1}{3} \delta^{\dot{a}\dot{b}}
f^{\dot{c}\dot{d}}_{(--)} \delta^{\dot{c}\dot{d}}.
\end{eqnarray*}
Then Eq. \eq{final-weyl} can be written as follows
\begin{equation} \la{f-weyl}
W_{ABCD}  = f^{ab}_{(++)}\eta^a_{AB} \eta^b_{CD} + f^{\dot{a}\dot{b}}_{(--)} \overline{\eta}^{\dot{a}}_{AB} \overline{\eta}^{\dot{b}}_{CD}
- \frac{1}{3} \big(f^{ab}_{(++)} \delta^{ab} +
f^{\dot{a}\dot{b}}_{(--)} \delta^{\dot{a}\dot{b}} \big)
(\delta_{AC} \delta_{BD}- \delta_{AD} \delta_{BC}).
\end{equation}
Combining the results in Eqs. \eq{dec-riemann} and \eq{f-weyl} gives us
the well-known decomposition of the curvature tensor $R$ into irreducible
components \ct{ahs-math,besse}, schematically given by
\begin{equation}\label{ahs-dec}
R = \left(
    \begin{array}{cc}
      W^+ + \frac{1}{12} s &  B  \\
      B^T &  W^- + \frac{1}{12} s \\
    \end{array}
  \right),
\end{equation}
where $s$ is the scalar curvature, $B$ is the traceless Ricci tensor, and
$W^\pm$ are the Weyl tensors.

One can similarly consider the self-duality equation for the Weyl tensor
that is defined by $W_{EFAB} = \pm \frac{1}{2} {\varepsilon_{AB}}^{CD} W_{EFCD}$ \ct{egh-report}.
An Einstein manifold is conformally self-dual if $g^{\dot{a}\dot{b}}_{(--)}=0$
and conformally anti-self-dual if $g^{ab}_{(++)} = 0$.
Note that the Weyl instanton (a conformally self-dual manifold) can also be regarded as
a Yang-Mills instanton and $\mathbb{C}P^2$ is a well-known example \ct{gibb-pope}.

In summary, we arrive at a remarkable result \ct{ohya} that any Einstein manifold with or without
a cosmological constant always arises as the sum of $SU(2)_L$ instantons
and $SU(2)_R$ anti-instantons. It explains why an Einstein manifold is stable
because two kinds of instantons belong to different gauge groups, one in $SU(2)_L$
and the other in $SU(2)_R$, and so they cannot decay into a vacuum.
The stability of an Einstein manifold will be further clarified in the part II \ct{partii}
by showing that the Einstein manifold carries nontrivial topological invariants.

\section{K\"ahler Manifolds and Twistor Space}

In this section we will survey some geometrical aspects of K\"ahler manifolds \ct{besse}
to illustrate the power of our gauge theory formulation.
Using the decomposition \eq{spin-sd-asd} of spin connections,
the torsion-free condition, $T^A =0$, in Eq. \eq{cartan-eq1} can equivalently be stated as the condition
that the triples in \eq{two-from} are covariantly constant \ct{jjl-hsy}, i.e.,
\begin{equation} \label{cov-comp-str}
D^{(+)} J^a_{+} \equiv d J^a_{+} - 2 \varepsilon^{abc} A^{(+)b} \wedge J^c_{+} = 0, \qquad
D^{(-)} J^{\dot{a}}_{-} \equiv d J^{\dot{a}}_{-} - 2 \varepsilon^{\dot{a}\dot{b}\dot{c}}
A^{(-)\dot{b}} \wedge J^{\dot{c}}_{-} = 0.
\end{equation}
$U(n)$ is the holonomy group of K\"ahler manifolds in $d=2n$-dimensions.
Therefore a four-dimensional K\"ahler manifold has $U(2)$ holonomy.
This means that the gauge group of spin connections for a K\"ahler manifold is
reduced from $SO(4) = SU(2)_L \times SU(2)_R$ to $U(2)$. The surviving $U(2)$ group depends on
the choice of K\"ahler form. To be specific, Eq. \eq{cov-comp-str} directly verifies that the K\"ahler condition, $d \Omega = 0$, for the K\"ahler form $\Omega = J^3_+$ can be satisfied with
$U(2) = U(1)_L \times SU(2)_R$ gauge fields by restricting $SO(4)$ gauge fields such that $A^{(+)1} = A^{(+)2} = 0$. And similarly the K\"ahler form $\Omega = J^3_-$ preserves $SU(2)_L \times U(1)_R$
gauge fields with $A^{(-)1} = A^{(-)2} =0$.
We may require a more stronger condition that one of the triples $(J^a_{+}, J^{\dot{a}}_{-})$
are entirely closed, for example, $dJ^{\dot{a}}_- = 0, \;\forall \dot{a}$.
This condition can be achieved by imposing $A^{(-)\dot{a}} = 0, \;\forall \dot{a}$
and so the manifold is half-flat, i.e. $F^{(-)\dot{a}} =0$, whose solution is called
a gravitational instanton \ct{egh-report}. In this case the manifold has $SU(2)_L$
(or $SU(2)_R$ for $dJ^{a}_+ = 0$) holonomy group.
Such a four-manifold is a hyper-K\"ahler manifold with $SU(2)$ holonomy which is also called
Calabi-Yau two-fold because it is Ricci-flat and K\"ahler \ct{besse}. An extra burden beyond
the hyper-K\"ahler condition makes a four-manifold be flat with trivial holonomy.

To be specific, suppose that $M$ is a complex manifold and
let us introduce local complex coordinates $z^\alpha =\{x^1 + i x^2, x^3 + i x^4 \},
\; \alpha=1,2$ and their complex conjugates $\bar{z}^{\bar{\alpha}}, \; \bar{\alpha}=1,2$,
in which an almost complex structure $J$ takes the form ${J^\alpha}_\beta
= i {\delta^\alpha}_\beta, \; {J^{\bar{\alpha}}}_{\bar{\beta}}
= - i {\delta^{\bar{\alpha}}}_{\bar{\beta}}$ \ct{naka-book}.
Note that, relative to the real basis $x^M, M=1, \cdots, 4$, the complex structure $J$
is given by $T^3_+ = i \tau^2 \otimes \mathbf{I}_2$ in Eq. \eq{thooft-matrix}.
One may choose a different complex structure
where local complex coordinates are given by $z^\alpha =\{x^1 + i x^2, x^3 - i x^4 \}$.
In this case the almost complex structure takes the form $J' = T^3_-
= i \tau^2 \otimes \tau^3$ which is related to $J$ by a parity transformation $P: x^4 \to - x^4$,
i.e., $J' = P J P$. And they commute each other, $JJ' = J'J$.
Therefore there are two independent K\"ahler manifolds defined by the complex structures $J$ and $J'$.
The decomposition \eq{hodge-2f} suggests that each K\"ahler structure is associated with
an instanton or an anti-instanton.

Let us further impose Hermitian condition on the complex manifold $M$ defined
by $g(X,Y) = g (JX,JY)$ for any $X,Y \in TM$.
This means that the Riemannian metric $g$ on a complex manifold $M$ is a Hermitian metric,
i.e., $g_{\alpha\beta} = g_{\bar{\alpha}\bar{\beta}} = 0, \; g_{\alpha\bar{\beta}}
= g_{\bar{\beta}\alpha}$ \ct{naka-book}.
The Hermitian condition can be solved by taking the vierbeins as
\begin{equation}\label{holo-vielbein}
    E^i_{\bar{\alpha}} = E^{\bar{i}}_{\alpha} = 0 \quad {\rm or} \quad
    E_i^{\bar{\alpha}} = E_{\bar{i}}^{\alpha} = 0
\end{equation}
where a tangent space index $A=1,\cdots,4$ has been split into a holomorphic index $i=1,2$ and
an anti-holomorphic index $\bar{i}=1,2$. This in turn means that
${J^i}_j = i {\delta^i}_j, \; {J^{\bar{i}}}_{\bar{j}} = - i {\delta^{\bar{i}}}_{\bar{j}}$.
Then one can see that the two-form $\Omega = J^3_+$ is a K\"ahler form with
respect to the complex structure $J$, i.e., $\Omega(X,Y) = g(JX,Y)$ and
similarly $\Omega(X,Y) = g(J'X,Y)$ for $\Omega = J^3_-$. And it is given by
\begin{equation}\label{kahler}
    \Omega = \frac{i}{2} E^i \wedge E^{\bar{i}} = \frac{i}{2} E^{i}_{\alpha}
E^{\bar{i}}_{\bar{\beta}}
    dz^\alpha \wedge d\bar{z}^{\bar{\beta}} =
    \frac{i}{2} g_{\alpha \bar{\beta}} dz^\alpha \wedge d\bar{z}^{\bar{\beta}}
\end{equation}
where $E^i = E^i_\alpha dz^\alpha$ is a holomorphic one-form and
$E^{\bar{i}} = E^{\bar{i}}_{\bar{\alpha}} d\bar{z}^{\bar{\alpha}}$ is
an anti-holomorphic one-form. The condition for a Hermitian manifold $(M, g)$ to be
K\"ahler is given by $d \Omega = 0$ for the K\"ahler form $\Omega = J^3_\pm$.
From Eq. \eq{cov-comp-str}, one can see that the K\"ahler condition leads to $U(2)$ gauge fields
such that $A^{(\pm)1} = A^{(\pm)2} = 0$ and thus $F^{(\pm)1} = F^{(\pm)2} = 0$.
In other words, the spin connections are $U(2)$-valued, i.e.,
\begin{equation}\label{spin-u2}
    \omega_{ij} = \omega_{\bar{i}\bar{j}} = 0,
\end{equation}
which immediately follows from Eq. \eq{su2-lr-field} using Eqs. \eq{complex-thooft} and \eq{anti-complex-thooft}. Hence, one can read off from Eq. \eq{einstein-mfd} that,
for a K\"ahler manifold $M$, $f^{ab}_{(\pm\pm)} = 0$ except $f^{33}_{(\pm\pm)} \neq 0$
and so $U(1)_{L}$ or $U(1)_R$ field strength among the $U(2)$ gauge fields is given by
\begin{equation}\label{ricci-form}
F^{(\pm)3} = dA^{(\pm)3} = f^{33}_{(\pm\pm)} J^3_\pm = f^{33}_{(\pm\pm)} \Omega.
\end{equation}
It is well-known \ct{besse,naka-book} that the Ricci tensor of a K\"ahler manifold $M$
is the field strength of the $U(1)$ part of spin connections.
It is obvious from Eq. \eq{dec-ricci} that
the Ricci tensor is given by $R_{AB} = 2 f^{33}_{(\pm\pm)} \delta_{AB}$ and
so $F^{(\pm)3}$ is a Ricci form of the K\"ahler manifold $M$
which defines the first Chern class $c_1(M) \equiv [F^{(\pm)3}/\pi] \in H^2(M, \mathbb{R})$.
Therefore one can see that the complex structures $J$ and $J'$ introduced above correspond
to the $U(1)$ generators $T^3_+$ and $T^3_-$, respectively, whose field strengths are
given by the Ricci form \eq{ricci-form} and define $U(1)$ (anti-)instantons of a K\"ahler manifold.
The result \eq{ricci-form} will be useful later to prove
some identity for topological invariants \ct{partii}.

A K\"ahler manifold $M$ with vanishing first Chern class, $c_1(M)=0$,
is called a Calabi-Yau manifold \ct{besse}.
Then the Calabi-Yau manifold in four dimensions is described by the Riemann curvature tensor
in Eq. \eq{einstein-mfd} with the coefficients satisfying $f^{\dot{a}\dot{b}}_{(--)} = 0$ (self-dual)
or $f^{ab}_{(++)} = 0$ (anti-self-dual) \ct{ohya}. In other words, the Riemann curvature tensor
obeys the self-duality relation defined by \ct{g-instanton}
\begin{equation}\label{sd-riemann}
    R_{ABEF} = \pm \frac{1}{2} {\varepsilon_{AB}}^{CD} R_{CDEF}
\end{equation}
and such a self-dual manifold is called a gravitational (anti-)instanton.
That is, gravitational instantons are half-flat, i.e., $F^{(+)a} = 0$ or $F^{(-)\dot{a}} = 0$,
and so one can always choose a self-dual gauge $A^{(+)a} = 0$ or $A^{(-)\dot{a}} = 0$,
respectively \ct{egh-report}. Then Eq. \eq{cov-comp-str} implies that there exists
a triple of K\"ahler forms,
to say $dJ^a_+ = 0$ or $dJ^{\dot{a}}_- = 0$. To recapitulate, a four-manifold $M$ satisfying
the self-duality in Eq. \eq{sd-riemann} is a hyper-K\"ahler manifold or equivalently
Ricci-flat and K\"ahler. Since the holonomy group
of a hyper-K\"ahler manifold is $SU(2) \cong Sp(1)$ which is a normal subgroup of $SO(4)$,
it follows that a hyper-K\"ahler manifold is simultaneously K\"ahler relative
to the triple $(I,J,K)$ of complex structures \ct{besse}.
This triple $(I,J,K)$ can be identified with the $SU(2)$ generators $T^a_+$
or $T^{\dot{a}}_-$ in \eq{thooft-matrix} which belong to another normal subgroup
of $SO(4)$ seeing zero curvature \ct{jjl-hsy}.
In fact the hyper-K\"ahler manifold has a continuous family of
K\"ahler structures defined by $aI + bJ + cK$ where $(a,b,c) \in \mathbb{S}^2$,
and this leads to the twistor theory of hyper-K\"ahler manifolds
\ct{t-penrose,twistor-book}.

The twistor space $\mathcal{Z}$ of a hyper-K\"ahler manifold $M$ is the product of
$M$ with two-sphere, i.e., $\mathcal{Z} = M \times \mathbb{S}^2$ where the two-sphere
parameterizes the complex structures of $M$ \ct{t-penrose}.
A choice of projective coordinates in $\mathbb{C}P^1 = \mathbb{S}^2$ corresponds to a choice
of a preferred complex structure, e.g., $J$. Therefore the twistor space $\mathcal{Z}$
can be viewed as a fiber bundle over $\mathbb{S}^2$ with a hyper-K\"ahler manifold $M$ as a fiber.
Let $(\omega_1, \omega_2, \omega_3)$ be the K\"ahler forms corresponding to $(I,J,K)$
on a hyper-K\"ahler manifold $M$, which can be identified with one of the triples in Eq. \eq{two-from}.
If we fix one of the K\"ahler structures, say $J = T^3_+$
or $T^3_-$ with the K\"ahler form $\omega_3 = \Omega$, then the two-form $\Phi \equiv
\frac{1}{2}(\omega_1 + i \omega_2) = -\frac{i}{2} E^1 \wedge E^2$ is of type (2,0) and determines a holomorphic symplectic structure. Eq. \eq{vol-id} then leads to the relation
\begin{equation}\label{vol-holo}
    2 \Phi \wedge \overline{\Phi} = \Omega \wedge \Omega.
\end{equation}
On a local chart, one can choose local Darboux coordinates $(z^1,z^2)$ for the (2,0)-form
$\Phi$ such that $\Phi = -\frac{i}{2} dz^1 \wedge dz^2$.
Let us consider a deformation of the holomorphic (2,0)-form $\Phi$ as follows
\be \la{deformation}
\Psi (t) = \Phi + i t \Omega + t^2 \overline{\Phi}
\ee
where the parameter $t$ takes values in $\mathbb{C}P^1=\mathbb{S}^2$.
One can easily see that $d\Psi(t) = 0$ for a hyper-K\"ahler manifold $M$ and
\be \la{psi-psi}
\Psi (t) \wedge \Psi (t) = 0
\ee
by Eq. \eq{vol-holo}. Since the two-form $\Psi(t)$ is closed and degenerate,
one can solve Eq. \eq{psi-psi} by introducing a $t$-dependent map $(z^1,z^2)
\to (Z^1(t;z^\alpha, \bar{z}^{\bar{\alpha}}), Z^2(t;z^\alpha, \bar{z}^{\bar{\alpha}}))$
such that \ct{oogu-vafa}
\be \la{symplectic-form}
\Psi (t) = - \frac{i}{2}dZ^1(t;z^\alpha, \bar{z}^{\bar{\alpha}}) \wedge
dZ^2(t;z^\alpha, \bar{z}^{\bar{\alpha}})
\ee
where the exterior derivative acts only along $M$ and not along $\mathbb{C}P^1$.
The $t$-dependent coordinates $Z^\alpha(t;z,\zbar)$ correspond to holomorphic (Darboux)
coordinates on a local chart where the 2-form $\Psi(t)$ becomes the holomorphic (2,0)-form.

When $t$ is small, one can solve \eq{symplectic-form} by expanding
$Z^\alpha(t;z,\zbar)$ in powers of $t$ as
\be \la{small-expansion}
Z^\alpha(t;z,\zbar) = z^\alpha + \sum_{n=1}^{\infty} \frac{t^n}{n}p_n^\alpha(z,\zbar).
\ee
By substituting this into Eq.\eq{deformation}, one gets at ${\cal O}(t)$
\bea \la{exp-eq1}
&& \partial_{\alpha} p_1^\alpha = 0, \\
\la{exp-eq2}
&& \Omega = - \frac{1}{2} \epsilon_{\alpha\beta} \bar{\partial}_{\bar{\gamma}} p_1^\beta
dz^\alpha \wedge d\bar{z}^{\bar{\gamma}}
\eea
where the fact was used that $\Omega$ is a (1,1)-form.
Eq. \eq{exp-eq1} can be solved by setting $p_1^\alpha = i \epsilon^{\alpha\beta} \partial_{\beta}
K$ and then $\Omega = i/2  \partial_\alpha \bar{\partial}_{\bar{\beta}} K dz^\alpha \wedge
d\bar{z}^{\bar{\beta}}$. From Eq. \eq{kahler}, one can identify the K\"ahler metric as
$g_{\alpha\bar{\beta}} = \partial_\alpha \bar{\partial}_{\bar{\beta}} K$ where
the real-valued smooth function $K(z,\zbar)$ is called the K\"ahler potential.
In terms of this K\"ahler potential $K$, Eq. \eq{psi-psi} can be written
as the complex Monge-Amp\`ere equation defined
by $\det(\partial_\alpha {\bar \partial_{\bar{\beta}}} K) = 1$ \ct{oogu-vafa}.

When $t$ is large, one can introduce another Darboux coordinates
${\widetilde Z}^{\bar{\alpha}}(t;z, \bar{z})$ such that
\be \la{symplectic-form2}
\Psi (t) = \frac{it^2}{2} d{\widetilde Z}^1(t;z^\alpha, \bar{z}^{\bar{\alpha}}) \wedge
d{\widetilde Z}^2(t;z^\alpha, \bar{z}^{\bar{\alpha}})
\ee
with expansion
\be \la{small-expansion2}
{\widetilde Z}^{\bar{\alpha}}(t;z,\zbar) = \zbar^{\bar{\alpha}} + \sum_{n=1}^{\infty} \frac{(-t^{-1})^n}{n}
{\widetilde p}_n^{\bar{\alpha}}(z,\zbar).
\ee
One can get the solution \eq{deformation} with ${\widetilde p}_1^{\bar{\alpha}} = - i
\epsilon^{\bar{\alpha}\bar{\beta}} \bar{\partial}_{\bar{\beta}} K$ and
$\Omega = i/2 \partial_\alpha \bar{\partial}_{\bar{\beta}} K dz^\alpha \wedge
d\bar{z}^{\bar{\beta}}$.

Let us introduce the real structure $\mathfrak{R}$ on $\mathbb{C}P^1$ defined
by complex conjugation composed with the antipodal map, e.g., $\mathfrak{R}
[Z^\alpha(t)] = {\bar Z}^{\bar{\alpha}}(-\frac{1}{t})
= {\widetilde Z}^{\bar{\alpha}}(t)$ \ct{lind-roc}.
From Eq. \eq{vol-holo}, we see that the two-form $\Psi(t)$ obeys the reality condition
\begin{equation}\label{reality}
   \Psi(t) = t^2  \mathfrak{R} [\Psi(t)]
\end{equation}
and so we have
\begin{eqnarray}\label{reality-eq}
   - \frac{i}{2}dZ^1(t) \wedge dZ^2(t) &=& \frac{it^2}{2} d{\bar Z}^1 \Big(-\frac{1}{t} \Big) \wedge
d{\bar Z}^2 \Big(-\frac{1}{t} \Big) \xx
&=& \frac{it^2}{2} d\widetilde{Z}^1 (t) \wedge
d\widetilde{Z}^2 (t).
\end{eqnarray}
The above reality relation shows that $Z^\alpha$ are related to $\bar{Z}^{\bar{a}}$
by a symplectomorphism up to the $t^2$-factor. We introduce a generating function $f(t; Z^1, \bar{Z}^1)$
for this twisted symplectomorphism defined by \ct{lind-roc}
\begin{equation}\label{sgen-func}
    Z^2 = - t \frac{\partial f}{\partial Z^1}, \qquad
\bar{Z}^2 = - \frac{1}{t} \frac{\partial f}{\partial \bar{Z}^1}
\end{equation}
and then
\begin{equation}\label{reality-twistor}
   - \frac{i}{2}dZ^1(t) \wedge dZ^2(t) = \frac{it}{2}
\frac{\partial^2 f}{\partial Z^1 \partial \bar{Z}^1} dZ^1 \wedge d{\bar Z}^1
\equiv \frac{it}{2} \partial \bar{\partial} f,
\end{equation}
where $\partial$ and $\bar{\partial}$ are holomorphic and anti-holomorphic exterior derivatives,
respectively, with respect to a complex structure $J$ at the north pole of $\mathbb{C}P^1$
and again act only on $M$ and not along the $\mathbb{C}P^1$. Since $\Psi(t)$ is a globally
defined holomorphic two-form, Eq. \eq{sgen-func} implies that $t \frac{\partial f}{\partial Z^1}$
is regular at the north pole and, hence, for a contour encircling $t=0$,
\begin{equation}\label{twistor-contour}
  \oint \frac{dt}{2\pi i}  t^n \frac{\partial f}{\partial Z^1} = 0, \qquad {n \geq 1}.
\end{equation}

Thus the function $f(t; Z^1, \bar{Z}^1)$ plays the role of a generating function for
symplectomorphisms between south and north poles.
In this way, the complex geometry of the twistor space $\mathcal{Z}$ encodes all
the information about the K\"ahler geometry of self-dual 4-manifolds \ct{twistor-book}.
We note that the exactly same construction of the twistor space $\mathcal{Z}$ can be applied
to noncommutative $U(1)$ instantons \ct{hsy-epl} which were proven
to be equivalent to gravitational instantons \ct{hsy12,hsy-jhep}.
We will further explore in part II \ct{partii} (a sequel of the present work)
the complex geometry of the twistor space $\mathcal{Z}$ and
its possible implications for spacetime foams.

\section{Four-Manifolds with Matter Coupling}

Our formalism can be fruitfully applied to the deformation theory of
Einstein spaces. First of all, it will be interesting to see how the
energy-momentum tensor $T_{AB}$ of matter fields in the Einstein equation
\begin{equation}\label{einstein-eq}
G_{AB} + \Lambda \delta_{AB} = 8 \pi G T_{AB}
\end{equation}
deforms the structure of an Einstein manifold described by Eq. \eq{einstein-mfd}.
First note that, among the 20 components of Riemann curvature tensor, the half of them describes gravitational degrees of freedom related to the Weyl tensor and the other half describes
matter degrees of freedom related to the Ricci tensor.
The Weyl tensor \eq{weyl} is a part of the curvature of spacetime that is not locally
determined by the matter through the Einstein equations \ct{st-book}.
Therefore, the deformation of an Einstein manifold by a coupling of matter fields affects only
the Ricci tensor part while keeping the Weyl tensor intact.
To see this, let us decompose the energy-momentum tensor $T_{AB}$
into a traceless part and a trace part as follow
\begin{eqnarray}\label{em-decom}
    T_{AB} &=& T_{AB} - \frac{1}{4} \delta_{AB} T + \frac{1}{4} \delta_{AB} T \xx
&\equiv& \widetilde{T}_{AB} + \frac{1}{4} \delta_{AB} T
\end{eqnarray}
where $T = T_{AA}$. By comparing Eq. \eq{dec-einstein} with Eq. \eq{em-decom},
one can deduce the following general result
\begin{eqnarray} \la{g-eins-1}
&& f^{a\dot{b}}_{(+-)}\eta^a_{AC} \overline{\eta}^{\dot{b}}_{BC} = 4 \pi G \widetilde{T}_{AB}, \\
\la{g-eins-2}
&&  f^{ab}_{(++)} \delta^{ab} =  f^{\dot{a}\dot{b}}_{(--)} \delta^{\dot{a}\dot{b}} =
\frac{\Lambda}{2} - \pi G T.
\end{eqnarray}
Motivated by the relation \eq{g-eins-1}, one may expand the traceless energy-momentum
tensor $\widetilde{T}_{AB}$ as
\begin{equation}\label{tr0-em}
    \widetilde{T}_{AB} = t^{a\dot{b}}_{(+-)}\eta^a_{AC} \overline{\eta}^{\dot{b}}_{BC}.
\end{equation}
This expansion is consistent with the fact that $\widetilde{T}_{AB}$ is a symmetric,
traceless second-rank tensor and so has 9 components. In other words,
one can invert the expression \eq{tr0-em} as
\begin{equation}\label{i-tr0-em}
    t^{a\dot{b}}_{(+-)} = \frac{1}{4} \eta^a_{AC} \overline{\eta}^{\dot{b}}_{BC} \widetilde{T}_{AB}.
\end{equation}
Then Eq. \eq{g-eins-1} reduces to a simple relation $f^{a\dot{b}}_{(+-)} = 4 \pi G t^{a\dot{b}}_{(+-)}$.

From the irreducible decomposition \eq{ahs-dec} of curvature tensor, we know that
the components $f^{a\dot{b}}_{(+-)}$ describe the traceless Ricci tensor denoted as $B$ and $B^T$
and $f^{ab}_{(++)} \delta^{ab} =  f^{\dot{a}\dot{b}}_{(++)} \delta^{\dot{a}\dot{b}}$ is
the Ricci scalar part denoted as $s$.
One can then draw a general conclusion from Eqs. \eq{g-eins-1} and \eq{g-eins-2}
even before considering a specific matter coupling. First of all,
the Einstein equations written in the form of Eqs. \eq{g-eins-1} and \eq{g-eins-2}
show us a crystal-clear picture how matter fields deform the structure of an Einstein manifold.
They in general introduce a mixing between $SU(2)_L$ and $SU(2)_R$ sectors, i.e.,
$f^{a\dot{b}}_{(+-)} \neq 0$. But, if $T=0$, such a matter field does not disturb
the conformal structure given by Eq. \eq{final-weyl} and the instanton structure
described by Eq. \eq{einstein-mfd}. This will be the case if matter fields preserve a
conformal symmetry and so their energy-momentum tensor is traceless.
We know that spin-one gauge fields in four-dimensions permit the conformal symmetry.
But other fields such as scalar and Dirac fields do not admit the conformal symmetry and
so they will also deform the instanton structure of an underlying Einstein manifold
through Eq. \eq{g-eins-2}.

To be specific, consider the Einstein theory coupled to matter fields where
the energy-momentum tensors of scalar fields, spinors and Yang-Mills gauge fields
are, respectively, given by
\begin{eqnarray} \label{scalar-em-tensor}
&& T^{(0)}_{AB} = E_A \phi^\mu E_B \phi^\mu - \delta_{AB} \mathcal{L}^{(0)}, \\
\la{dirac-em-tensor}
&& T^{(1/2)}_{AB} =\frac{1}{2} (\overline{\psi} \Gamma_A D_B \psi +
\overline{\psi} \Gamma_B D_A \psi) - \delta_{AB} \mathcal{L}^{(1/2)}, \\
\label{ym-em-tensor}
&& T^{(1)}_{AB} = \frac{2}{g^2_{YM}} \mathrm{Tr} \Big(F_{AC} F_{BC}
- \frac{1}{4} \delta_{AB} F_{CD} F^{CD} \Big),
\end{eqnarray}
where $E_A \phi^\mu = E^M_A \partial_M \phi^\mu \;(\mu = 1, \cdots, n)$ and $\mathcal{L}^{(0)} = \frac{1}{2} g^{MN} \partial_M \phi^\mu \partial_N \phi^\mu - V(\phi^\mu)$ and
$D_A \psi = (E_A + \omega_A) \psi$ and
$\mathcal{L}^{(1/2)} = \overline{\psi} \Gamma^A D_A \psi - V(\overline{\psi}, \psi)$.
In Euclidean space, the Dirac spinor $\psi$ has four complex components and the conjugate
spinor is defined by $\overline{\psi} = \psi^\dagger \Gamma^5$ and the Majorana spinor
is a bit more subtle to define. We refer to \ct{NW} for Euclidean spinors.
From the above results, one can see that only $T^{(1)}_{AB}$ is traceless and
so Yang-Mills gauge fields do not deform Eq. \eq{g-eins-2} but affect only Eq. \eq{g-eins-1}.
Of course, this is a consequence of the conformal symmetry of Yang-Mills gauge theory.

The Yang-Mills field strength $F_{AB}$ in the adjoint representation of gauge group $G$
can also be decomposed like \eq{dec-su2l} or \eq{dec-su2r} according to the Hodge
decomposition \eq{dec-2f}:
\begin{equation}\label{dec-ym}
    F_{AB} \equiv f^{a}_{(+)}\eta^a_{AB} + f^{\dot{a}}_{(-)}
\overline{\eta}^{\dot{a}}_{AB}.
\end{equation}
It is then straightforward to calculate the energy-momentum tensor \eq{ym-em-tensor}
which is given by \ct{ohya}
\begin{equation}\label{dec-em-tensor}
\widetilde{T}^{(1)}_{AB} = \frac{4}{g^2_{YM}} \mathrm{Tr} \big( f^{a}_{(+)} f^{\dot{b}}_{(-)} \big)
\eta^a_{AC} \overline{\eta}^{\dot{b}}_{BC}
\end{equation}
or
\begin{equation}\label{t-ym}
  t^{a\dot{b}}_{(+-)} =  \frac{4}{g^2_{YM}} \mathrm{Tr} \big( f^{a}_{(+)} f^{\dot{b}}_{(-)} \big)
\end{equation}
and $T^{(1)} = 0$. Thus Eq. \eq{g-eins-2} is not deformed by Yang-Mills gauge fields
as a result of the conformal symmetry and substituting Eq. \eq{dec-em-tensor} into Eq. \eq{g-eins-1}
leads to the deformed equations instead of Eq. \eq{cond-einstein}
\begin{eqnarray}\label{def-einstein}
  && f^{ab}_{(++)} \delta^{ab} = f^{\dot{a}\dot{b}}_{(--)} \delta^{\dot{a}\dot{b}}
  = \frac{\Lambda}{2}, \xx
 && f^{a\dot{b}}_{(+-)} = \frac{16 \pi G}{g^2_{YM}}
  \mathrm{Tr} \big( f^{a}_{(+)} f^{\dot{b}}_{(-)} \big).
\end{eqnarray}

It is straightforward to determine the mixing coefficients $f^{a\dot{b}}_{(+-)}$ for scalar
and spinor fields by calculating the energy-momentum tensor in Eq. \eq{i-tr0-em}
to which any terms proportional to $\delta_{AB}$ do not contribute thanks to the property
$\eta^a_{AB} \overline{\eta}^{\dot{b}}_{AB} = 0$. Also the correction of the Ricci scalar part
can be calculated by Eq. \eq{g-eins-2}. But note that this modification of the Ricci scalar part
will also affect the Weyl tensor part through the structure \eq{f-weyl}.
It should be the case because the scalar and spinor fields do not respect the conformal symmetry
and so the Weyl tensor will be corrected by the presence of these fields.\footnote{It is
interesting to notice that the traceless Ricci tensor and the Ricci scalar belong
to completely different blocks as shown up in Eq. \eq{ahs-dec} although the Ricci scalar
is defined as the trace of the Ricci tensor. The Ricci scalar rather belongs to the same block
as the Weyl tensor.} In conclusion scalar and spinor fields introduce a mixing between
self-dual and anti-self-dual sectors of curvature tensors to deform the underlying structure
of an Einstein manifold as the manner described by Eqs. \eq{g-eins-1} and \eq{g-eins-2}.

\section{Discussion}

We would like to emphasize that the Lemma proven in Section 4 holds not only for 4-dimensional
spin manifolds but also for general oriented 4-manifolds although we have introduced a spinor
representation of $SO(4)$ to prove it. Actually we need only two ingredients to prove the Lemma,
as we briefly outlined in the Introduction. Recall that if $M$ is an oriented 4-manifold,
the structure group of $TM$, a tangent bundle over $M$, is $SO(4)$ whose Lie algebra
is isomorphic to $SU(2)_L \times SU(2)_R$ and the Hodge $*$-operation is an involution
of the space $\Lambda^2 T^*M$ of two-forms which decomposes the two-forms into self-dual
and anti-self dual parts, both of which do not necessarily require
a spin structure of 4-manifold \ct{besse}.
Then the Clifford map \eq{clifford-map} introduces an isomorphic correspondence
between the splitting of $SO(4)$ and the Hodge decomposition:
\begin{equation}\label{clifford-correspondence}
    J^{AB}_\pm \equiv \frac{1}{2} (1 \pm \Gamma^5) J^{AB} \qquad
    \Leftrightarrow \qquad F^{(\pm)} = \frac{1}{2}(1 \pm * )F
\end{equation}
where both $\frac{1}{2} (1 \pm \Gamma^5)$ and $\frac{1}{2}(1 \pm * )$ are projection operators
acting on the $SO(4)$ Lie algebra and $\Lambda^2 T^*M$, respectively. See Eq. \eq{hodge-2f}.
These two are enough to derive the Lemma. For example, though $\mathbb{C}P^2$ admits only
a generalized spin structure, $Spin^{\mathbb{C}}$-structure,
one can get the decomposition \eq{dec-riemann} with impunity \ct{ohya}.

In the Donaldson's theory of 4-manifolds \ct{4man-book}, Yang-Mills theory shows a profound
play in describing the global structure of 4-manifolds where the moduli space
of (gauge-inequivalent) solutions to the self-dual Yang-Millls equations plays the central role.
Let us survey the Lemma again
to get some insight about the Donaldson's theory. Suppose that $M$ is an Einstein manifold
such that it admits a metric $g$ obeying Eq. \eq{einstein-equation}. Given such a metric $g$,
one can continuously perturb to a new metric $g + \delta g$ such that it still describes an Einstein
manifold obeying Eq. \eq{einstein-equation}. Following the identification \eq{spin-sd-asd},
we can translate the metric perturbation as the perturbation of $SU(2)$ gauge fields
$A_M^{(\pm)}$, i.e., $A_M^{(\pm)} \to A_M^{(\pm)} + \delta A_M^{(\pm)}$. The Lemma then implies that
the Einstein condition for the perturbed metric can be interpreted as instanton connections
for the $SU(2)$ gauge fields $A_M^{(\pm)} + \delta A_M^{(\pm)}$ satisfying Eq. \eq{ym-instanton-eq}
from the gauge theory point of view. Hence the perturbed connections $\delta A_M^{(\pm)}$ will take values in the moduli space of $SU(2)$ Yang-Mills instantons over an Einstein manifold $M$ \ct{4man-book}.
However the variational problem for Eq. \eq{ym-instanton-eq} is more complicated
than that for usual instantons in a fixed
background because the four-dimensional metric used to define Eq. \eq{ym-instanton-eq}
simultaneously determines $SU(2)$ instanton connections too. It may be more transparent
by writing Eq. \eq{ym-instanton-eq} as the form \ct{opy}
\begin{equation}\label{su2-lo-instanton}
     F^{(\pm)}_{MN} = \pm \frac{1}{2}\frac{\varepsilon^{RSPQ}}{\sqrt{g}}
     g_{MR}g_{NS}  F^{(\pm)}_{PQ}
\end{equation}
where $\sqrt{g} = \det E_M^A$ and $\varepsilon^{MNPQ}$ is the metric independent Levi-Civita symbol
with $\varepsilon^{1234} = 1$. Therefore it is necessary
to consider the variations $\delta g$ as well as $\delta A_M^{(\pm)}$ in Eq. \eq{su2-lo-instanton}
to define a deformation complex for the Einstein structures on $M$.
However, it may be worthwhile to retain the fact that the variations $\delta g$ and $\delta A_M^{(\pm)}$
are not independent but related to each other by Eq. \eq{spin-sd-asd}.
All in all, the moduli space of Einstein metrics seems to be essentially the tensor product
of the moduli spaces of self-dual and anti-self-dual instantons whose connections are defined
by Eq. \eq{spin-sd-asd} in terms of the spin connections of the Einstein metric itself.
The simplest case to test the conjecture is to consider the moduli space of hyper-K\"ahler
(or half-flat) structures satisfying Eq. \eq{sd-riemann} which would be given by only one of
the two factors since the other part just sees flat connections.
We hope to address this problem elsewhere.

Our gauge theory formulation of Einstein gravity has relied on the fact that
spin connections in the tetrad formalism are gauge fields of Lorentz group \ct{mtw-book}.
But the fundamental variables in the tetrad formalism are vierbeins $E^M_A(x)$ or
the orthonormal tangent vectors $E_A = E^M_A(x)\partial_M$ in Eq. \eq{dual-vector}
rather than the spin connections.
The spin connections are determined by the vierbeins as Eq. \eq{spin-conn}
via the torsion free condition. On the contrary, the gauge theory has no analogue
of vierbeins or a Riemannian metric, as we remarked in the footnote \ref{gravity-gauge}.
See the Table 1 in \ct{opy} for some crucial differences between gravity and gauge theory.
Therefore, the connection between gravity and gauge theory is still incomplete
although we could have understood the Einstein equation for four-manifolds
as the self-duality equation of Yang-Mills instantons. Is it possible to find a gauge theory
representation of gravity including Riemannian metrics ?

Now we will show that the vierbeins and so the Riemannian metrics arise from electromagnetic
fields living in a space $(M, B)$ supporting a symplectic structure $B$ \ct{hsy-jhep,em-gravity,hsy-siva}.\footnote{\label{symp-pois}
The symplectic structure $B$ is a nondegenerate, closed 2-form, i.e. $dB=0$ \ct{mechanics-book}.
Therefore the symplectic structure $B$ defines a bundle isomorphism $B: TM \to T^*M$ by $X \mapsto A
= \iota_X B$ where $\iota_X$ is an interior product with respect to a vector
field $X \in \Gamma(TM)$. One can invert this map to obtain the inverse map $\theta \equiv B^{-1}:
T^*M \to TM$ defined by $\alpha \mapsto X = \theta(\alpha)$ such that $X(\beta) =
\theta(\alpha,\beta)$ for $\alpha, \beta \in \Gamma(T^*M)$.
The bivector $\theta \in \Gamma(\Lambda^2 TM)$ is called a Poisson structure of $M$
which defines a bilinear operation on $C^\infty(M)$, the so-called Poisson bracket,
defined by $\{f, g\}_\theta = \theta(df, dg)$ for $f, g \in C^\infty(M)$.
Then the real vector space $C^\infty(M)$, together with the
Poisson bracket $\{-, -\}_\theta$, forms an infinite-dimensional Lie algebra,
called a Poisson algebra $\mathfrak{P} =(C^\infty(M), \{-, -\}_\theta)$.}
Recently the emergent gravity scheme based on large $N$ matrix models and noncommutative
field theories has drawn a lot of attention (see \ct{jjl-hsy,hsy-stein} for a review
of this subject and references therein). The emergent gravity scheme seems to grant
a radically new picture about gravity and provide a clue to realize a gauge theory
representation of gravity including Riemannian metrics.

First note that the orthonormal tangent vectors $E_A = E^M_A(x)\partial_M \in \Gamma(TM)$
satisfy the Lie algebra \eq{lie-vector}. In general, the composition $[X,Y]$, the Lie bracket
of $X$ and $Y$, on $\Gamma(TM)$, together with the real vector space structure of $\Gamma(TM)$,
forms a Lie algebra $\mathfrak{V} = (\Gamma(TM), [-,-])$. There is a natural Lie algebra
homomorphism between the Lie algebra $\mathfrak{V} = (\Gamma(TM), [-,-])$ and
the Poisson algebra $\mathfrak{P} =(C^\infty(M), \{-, -\}_\theta)$
(see the footnote \ref{symp-pois}) defined by \ct{mechanics-book}
\begin{equation}\label{lie-homo}
  C^\infty(M) \to  \Gamma(TM) : f \mapsto X_f
\end{equation}
such that
\begin{equation}\label{ham-vec}
    X_f (g) = - \theta(df, dg) = \{g, f\}_\theta
\end{equation}
for $f, g \in C^\infty(M)$. It is easy to prove the Lie algebra homomorphism
\begin{equation} \label{vec-homo}
X_{\{f, g\}_\theta} = - [X_f, X_g]
\end{equation}
using the Jacobi identity of the Poisson algebra $\mathfrak{P}$.

Let us take $M = \mathbb{R}^4$ and a constant symplectic structure $B = \frac{1}{2}
B_{MN} dx^M \wedge dx^N$, for simplicity. A remarkable point is that the electromagnetism
on a symplectic manifold $(M, B)$ is completely specified by the Poisson algebra
$\mathfrak{P} =(C^\infty(M), \{-, -\}_\theta)$ \ct{jjl-hsy}. For example, the action is given by
\begin{equation}\label{nc-action}
    S = \frac{1}{4 g_{YM}^2} \int d^4 x \{ D_A, D_B \}_\theta^2
\end{equation}
where
\begin{equation}\label{coordinate-d}
    D_A(x) = B_{AB} x^B + \widehat{A}_A(x) \in C^\infty(M), \qquad A = 1, \cdots, 4
\end{equation}
are covariant dynamical coordinates describing fluctuations from the Darboux coordinate $x^A$,
i.e. $\{x^A, x^B \}_\theta = \theta^{AB}$, and
\begin{eqnarray}\label{nc-field}
    \{ D_A(x), D_B(x) \}_\theta &=& -B_{AB} + \partial_A   \widehat{A}_B -
    \partial_B   \widehat{A}_A + \{ \widehat{A}_A, \widehat{A}_B \}_\theta \xx
    &\equiv&  -B_{AB} + \widehat{F}_{AB}(x) \in C^\infty(M).
\end{eqnarray}
It is clear that the equations of motion as well as the Bianchi identity can be represented
only with the Poisson bracket $\{-, -\}_\theta$.

A peculiar thing for the action \eq{nc-action} is that the field strength $\widehat{F}_{AB}$
in Eq. \eq{nc-field} is nonlinear due to the Poisson bracket term
although it is the curvature tensor of $U(1)$ gauge fields. Thus one can consider
a nontrivial solution of the following self-duality equation
\begin{equation}\label{u1-instanton}
    \widehat{F}_{AB} = \pm \frac{1}{2} {\varepsilon_{AB}}^{CD} \widehat{F}_{CD}.
\end{equation}
In fact, after the canonical Dirac quantization of the Poisson algebra $\mathfrak{P} =
(C^\infty(M), \{-, -\}_\theta)$, the solution of the self-duality equation \eq{u1-instanton}
is known as noncommutative $U(1)$ instantons \ct{nc-inst,hsy-ncinst}.
When applying the Lie algebra homomorphism \eq{vec-homo} to Eq. \eq{nc-field},
the self-duality equation \eq{u1-instanton} is mapped to the self-duality equation
of the vector fields $V_A \equiv X_{D_A} \in \Gamma(TM)$ obtained by the map \eq{ham-vec}
from the set of the covariant coordinates $D_A(x)$ in Eq. \eq{coordinate-d} \ct{hsy-epl,hsy-jhep}:
\begin{equation}\label{u1-graviton}
    [V_A, V_B] = \pm \frac{1}{2} {\varepsilon_{AB}}^{CD} [V_C, V_D].
\end{equation}
Note that the vector fields $V_A = V_A^M \partial_M$ are divergence free, i.e.,
$\partial_M V^M_A =0$ by the definition \eq{ham-vec} and so preserves a volume form $\nu$
because $\mathcal{L}_{V_A} \nu = (\nabla \cdot V_A) \nu = 0$
where $\mathcal{L}_{V_A}$ is a Lie derivative with respect to the vector field $V_A$.
Furthermore it can be shown \ct{hsy-jhep} that $V_A$ can be related to the vierbeins $E_A$ by $V_A
= \lambda E_A$ with $\lambda \in C^\infty(M)$ to be determined.

If the volume form $\nu$ is given by
\begin{equation}\label{volume}
    \nu \equiv \lambda^{-2} \nu_g = \lambda^{-2} E^1 \wedge \cdots \wedge E^4
\end{equation}
or, in other words, $\lambda^2 = \nu(V_1, \cdots, V_4)$,
one can easily check that the triple of K\"ahler forms in Eq. \eq{two-from}
is given by \ct{jjl-hsy}
\begin{equation}\label{ansatz}
    J_+^a = \frac{1}{2} \eta^a_{AB} \iota_A \iota_B \nu, \qquad
     J_-^{\dot{a}} = - \frac{1}{2} \overline{\eta}^{\dot{a}}_{AB} \iota_A \iota_B \nu,
\end{equation}
where $\iota_A$ is the interior product with respect to $V_A$.
In Section 5, we showed that gravitational instantons satisfying Eq. \eq{sd-riemann}
are hyper-K\"ahler manifolds, i.e., $dJ_+^a = 0$ or $dJ_-^{\dot{a}} = 0$ and vice versa.
It is straightforward to prove that the hyper-K\"ahler conditions
$dJ_+^a = 0$ or $dJ_-^{\dot{a}} = 0$ are precisely equivalent to Eq. \eq{u1-graviton}
which can easily be seen by applying to Eq. \eq{ansatz} the formula \ct{mechanics-book}
\begin{equation}\label{math}
    d(\iota_X \iota_Y \alpha) = \iota_{[X,Y]} \alpha + \iota_Y \mathcal{L}_X \alpha
    - \iota_X \mathcal{L}_Y \alpha + \iota_X \iota_Y d \alpha
\end{equation}
for vector fields $X, Y$ and a $p$-form $\alpha$.

In retrospect, Eq. \eq{u1-graviton} was derived from the self-duality equation \eq{u1-instanton}
of $U(1)$ gauge fields defined on the symplectic manifold $(\mathbb{R}^4, B)$.
As a consequence, $U(1)$ instantons on the symplectic manifold $(\mathbb{R}^4, B)$ are
gravitational instantons \ct{hsy-epl,hsy12,hsy-jhep} !
We want to emphasize that the emergence of Riemannian metrics from
symplectic $U(1)$ gauge fields is an inevitable consequence of the Lie algebra
homomorphism between the Poisson algebra $\mathfrak{P} =(C^\infty(M), \{-, -\}_\theta)$ and
the Lie algebra $\mathfrak{V} = (\Gamma(TM), [-,-])$ if the underlying action of
$U(1)$ gauge fields is given by the form of Eq. \eq{nc-action}. Moreover, the equivalence between
$U(1)$ instantons in the action \eq{nc-action} and gravitational instantons, as depicted in Figure 1,
turns out to be a particular case of more general duality between the $U(1)$
gauge theory on a symplectic manifold $(M,B)$ and Einstein gravity \ct{hsy-jhep,hsy-siva}.

A mysterious feature pops out when we add the relationship between noncommutative $U(1)$
instantons, Yang-Mills instantons and gravitational instantons altogether, as shown in Figure 1.
If the trinity relation in Figure 1 holds, there must be a relationship between
noncommutative $U(1)$ instantons and $SU(2)$ Yang-Mills instantons which is never explored so far.
This correspondence, if any, may debunk how $SU(2)$ gauge fields (in a intrepid term,
weak interaction) together with Einstein gravity arise from noncommutative $U(1)$ gauge fields.
We do not have any concrete understanding yet but it would be worthwhile to submit
the problem for a novel unification scheme.

In part II \cite{partii}, we will apply the gauge theory formulation of Euclidean gravity
to the topological classification of four-manifolds. There are two topological invariants
for a four-manifold $M$, namely the Euler characteristic $\chi(M)$ and the Hirzebruch
signature $\tau(M)$, which can be expressed as integrals of the curvature
of a four dimensional metric \ct{egh-report}.
The topological invariants of four-manifolds are basically characterized by configurations
of $SU(2)$ instantons and anti-instantons \ct{ohya}. We observe that the topological numbers of compact
Einstein manifolds appear on an even positive integer lattice and show an intriguing
reflection symmetry with respect to the interchange of $SU(2)$ instantons and anti-instantons,
which we call ``mirror" symmetry. The twistor space of hyper-K\"ahler manifolds discussed
in Section 5 will be further studied, especially, from the standpoint of the trinity relation
in Figure 1. It turns out that the decomposition of Riemann curvature tensors in Section 4 is
particularly useful for the Petrov and Bianchi classifications of Riemannian manifolds \ct{mtw-book}.
We will also study a general class of four-manifolds with vanishing Weyl curvature
with some cosmological implications \ct{st-book}.

\section*{Acknowledgments}

HSY thanks Sangheon Yun for helpful discussions.
This research was supported by Basic Science Research Program through the National Research
Foundation of Korea (NRF) funded by the Ministry of Education, Science and Technology (2011-0010597).
The work of H.S. Yang was also supported by the RP-Grant 2010 of Ewha Womans University.

\appendix

\section{'t Hooft symbols}

In this Appendix, we will not distinguish the two kinds of Lie algebra indices
$a \in SU(2)_L$ and $\dot{a} \in SU(2)_R$ for a notational simplicity (if necessary).
The 't Hooft symbols $\eta^a_{AB}$ and ${\overline \eta}^a_{AB}$ for $a = 1,2,3$
are defined by Eq. \eq{thooft-symbol} whose components can be explicitly determined by
\begin{eqnarray} \label{tHooft-symbol}
\begin{array}{l}
{\eta}^a_{AB} = {\varepsilon}^{a4AB} + \delta^{aA}\delta^{4B}
- \delta^{aB}\delta^{4A}, \\
{\overline \eta}^{a}_{AB} = {\varepsilon}^{a4AB} - \delta^{aA}\delta^{4B}
+ \delta^{aB}\delta^{4A}
\end{array}
\end{eqnarray}
with ${\varepsilon}^{1234} = 1$. Using the explicit result, it is straightforward to derive
the following identities for the 't Hooft symbols \cite{opy}
\begin{eqnarray} \label{self-eta}
&& \eta^{(\pm)a}_{AB} = \pm \frac{1}{2} {\varepsilon_{AB}}^{CD}
\eta^{(\pm)a}_{CD}, \\
\label{proj-eta}
&& \eta^{(\pm)a}_{AB}\eta^{(\pm)a}_{CD} =
\delta_{AC}\delta_{BD}
-\delta_{AD}\delta_{BC} \pm
\varepsilon_{ABCD}, \\
\label{self-eigen}
&& \varepsilon_{ABCD} \eta^{(\pm)a}_{DE} = \mp (
\delta_{EC} \eta^{(\pm)a}_{AB} + \delta_{EA} \eta^{(\pm)a}_{BC} -
\delta_{EB} \eta^{(\pm)a}_{AC} ), \\
\label{eta-etabar}
&& \eta^{(\pm)a}_{AB} \eta^{(\mp)b}_{AB}=0, \\
\label{eta^2}
&& \eta^{(\pm)a}_{AC}\eta^{(\pm)b}_{BC} =\delta^{ab}\delta_{AB} +
\varepsilon^{abc}\eta^{(\pm)c}_{AB}, \\
\label{eta-ex}
&& \eta^{(\pm)a}_{AC}\eta^{(\mp)b}_{BC} =
\eta^{(\mp)b}_{AC}\eta^{(\pm)a}_{BC}, \\
\label{eta-o4-algebra}
&& \varepsilon^{abc} \eta^{(\pm)b}_{AB} \eta^{(\pm)c}_{CD} =
    \delta_{AC} \eta^{(\pm)a}_{BD} - \delta_{AD} \eta^{(\pm)a}_{BC}
    - \delta_{BC} \eta^{(\pm)a}_{AD} + \delta_{BD} \eta^{(\pm)a}_{AC}
\end{eqnarray}
where $\eta^{(+)a}_{AB} \equiv \eta^a_{AB}$ and $\eta^{(-)a}_{AB} \equiv {\overline
\eta}^a_{AB}$.

If we introduce two families of $4 \times 4$ matrices defined by
\begin{equation} \label{thooft-matrix}
[T^a_+]_{AB} \equiv \eta^a_{AB}, \qquad [T^a_-]_{AB} \equiv {\overline
\eta}^a_{AB},
\end{equation}
the matrix representation in \eq{thooft-matrix} provides two independent spin $s=\frac{3}{2}$ representations of $SU(2)$ Lie algebra. Explicitly, they are given by
\begin{eqnarray}
&& T^{1}_+ =\begin{pmatrix}
      0 & 0 & 0 & 1 \\
      0 & 0 & 1 & 0 \\
      0 & -1 & 0 & 0 \\
      -1 & 0 & 0 & 0 \\
             \end{pmatrix}, \;\;
  T^{2}_+ = \begin{pmatrix}
      0 & 0 & -1 & 0 \\
      0 & 0 & 0 & 1 \\
      1 & 0 & 0 & 0 \\
      0 & -1 & 0 & 0 \\
    \end{pmatrix}, \;\;
  T^{3}_+ = \begin{pmatrix}
      0 & 1 & 0 & 0 \\
      -1 & 0 & 0 & 0 \\
      0 & 0 & 0 & 1 \\
      0 & 0 & -1 & 0 \\
    \end{pmatrix}, \\
&& T^{1}_- = \begin{pmatrix}
      0 & 0 & 0 & -1 \\
      0 & 0 & 1 & 0 \\
      0 & -1 & 0 & 0 \\
      1 & 0 & 0 & 0 \\
    \end{pmatrix}, \;\;
  T^{2}_- =  \begin{pmatrix}
      0 & 0 & -1 & 0 \\
      0 & 0 & 0 & -1 \\
      1 & 0 & 0 & 0 \\
      0 & 1 & 0 & 0 \\
    \end{pmatrix}, \;\;
  T^{3}_- =  \begin{pmatrix}
      0 & 1 & 0 & 0 \\
      -1 & 0 & 0 & 0 \\
      0 & 0 & 0 & -1 \\
      0 & 0 & 1 & 0 \\
    \end{pmatrix}
\end{eqnarray}
according to the definition \eq{tHooft-symbol}.
Indeed Eqs. (\ref{eta^2}) and (\ref{eta-ex}) immediately show that
$T^a_\pm$ satisfy $SU(2)$ Lie algebras, i.e.,
\begin{equation} \label{thooft-su2}
[T^a_\pm, T^b_\pm] = - 2 \varepsilon^{abc} T^c_\pm,
\qquad [T^a_\pm, T^b_\mp] = 0.
\end{equation}
The definition \eq{thooft-matrix} implies that the self-duality \eq{self-eta} is inherited to
the matrix representation
\begin{equation}\label{self-dual-su2}
   [T^a_\pm]_{AB} = \pm \frac{1}{2} {\varepsilon_{AB}}^{CD} [T^a_\pm]_{CD}.
\end{equation}

Finally we list the nonzero components of the 't Hooft symbols in the basis
of complex coordinates $z^\alpha = \{ z^1 = x^1 + i x^2, z^2 = x^3 + i x^4 \}$
and their complex conjugates $\bar{z}^{\bar{\alpha}}$:
\begin{equation} \la{complex-thooft}
\begin{array}{llll}
  \eta^1_{12} = -\frac{i}{2},  & \eta^2_{12} = -\frac{1}{2},
& \eta^3_{1\bar{1}} = \frac{i}{2}, & \eta^3_{2\bar{2}} = \frac{i}{2}
\end{array}
\end{equation}
where we denote $\eta^a_{\alpha\beta} = \eta^a_{z^\alpha z^\beta}, \; \eta^a_{\alpha\bar{\beta}}
= \eta^a_{z^\alpha \bar{z}^{\bar{\beta}}}$, etc. and the complex conjugates are not shown up
since they can easily be implemented. The corresponding values of $\overline{\eta}^{\dot{a}}_{AB}$
for the complex structure $J$ can be obtained from those of $\eta^a_{AB}$ by interchanging $z^2 \leftrightarrow \bar{z}^2$. But, with another complex structure $J'$ where complex coordinates
are given by $z^\alpha = \{ z^1 = x^1 + i x^2, z^2 = x^3 - i x^4 \}$, the nonzero components of $\overline{\eta}^{\dot{a}}_{AB}$ are the same as Eq. \eq{complex-thooft}:
\begin{equation} \la{anti-complex-thooft}
\begin{array}{llll}
  \overline{\eta}^1_{12} = -\frac{i}{2},  & \overline{\eta}^2_{12} = -\frac{1}{2},
& \overline{\eta}^3_{1\bar{1}} = \frac{i}{2}, & \overline{\eta}^3_{2\bar{2}} = \frac{i}{2}.
\end{array}
\end{equation}
The above result implies that the space of complex structure deformations for a given self-dual
structure can be identified with the homogeneous space $SO(4)/U(2) = \mathbb{C}P^1$.

\newpage

\nc{\PR}[3]{Phys. Rev. {\bf #1}, #2 (#3)}
\nc{\NPB}[3]{Nucl. Phys. {\bf B#1}, #2 (#3)}
\nc{\PLB}[3]{Phys. Lett. {\bf B#1}, #2 (#3)}
\nc{\PRD}[3]{Phys. Rev. {\bf D#1}, #2 (#3)}
\nc{\PRL}[3]{Phys. Rev. Lett. {\bf #1}, #2 (#3)}
\nc{\PREP}[3]{Phys. Rep. {\bf #1}, #2 (#3)}
\nc{\EPJ}[3]{Eur. Phys. J. {\bf C#1}, #2 (#3)}
\nc{\PTP}[3]{Prog. Theor. Phys. {\bf #1}, #2 (#3)}
\nc{\CMP}[3]{Commun. Math. Phys. {\bf #1}, #2 (#3)}
\nc{\MPLA}[3]{Mod. Phys. Lett. {\bf A#1}, #2 (#3)}
\nc{\CQG}[3]{Class. Quant. Grav. {\bf #1}, #2 (#3)}
\nc{\NCB}[3]{Nuovo Cimento {\bf B#1}, #2 (#3)}
\nc{\ANNP}[3]{Ann. Phys. (N.Y.) {\bf #1}, #2 (#3)}
\nc{\GRG}[3]{Gen. Rel. Grav. {\bf #1}, #2 (#3)}
\nc{\MNRAS}[3]{Mon. Not. Roy. Astron. Soc. {\bf #1}, #2 (#3)}
\nc{\JHEP}[3]{J. High Energy Phys. {\bf #1}, #2 (#3)}
\nc{\JCAP}[3]{JCAP {\bf #1}, #2 {#3}}
\nc{\ATMP}[3]{Adv. Theor. Math. Phys. {\bf #1}, #2 (#3)}
\nc{\AJP}[3]{Am. J. Phys. {\bf #1}, #2 (#3)}
\nc{\ibid}[3]{{\it ibid.} {\bf #1}, #2 (#3)}
\nc{\ZP}[3]{Z. Physik {\bf #1}, #2 (#3)}
\nc{\PRSL}[3]{Proc. Roy. Soc. Lond. {\bf A#1}, #2 (#3)}
\nc{\LMP}[3]{Lett. Math. Phys. {\bf #1}, #2 (#3)}
\nc{\AM}[3]{Ann. Math. {\bf #1}, #2 (#3)}
\nc{\hepth}[1]{{\tt [arXiv:hep-th/{#1}]}}
\nc{\grqc}[1]{{\tt [arXiv:gr-qc/{#1}]}}
\nc{\astro}[1]{{\tt [arXiv:astro-ph/{#1}]}}
\nc{\hepph}[1]{{\tt [arXiv:hep-ph/{#1}]}}
\nc{\phys}[1]{{\tt [arXiv:physics/{#1}]}}
\nc{\arxiv}[1]{{\tt [arXiv:{#1}]}}


\end{document}